\documentclass[lettersize,journal]{IEEEtran}
\usepackage{amsmath,amsfonts}
\usepackage[ruled,vlined]{algorithm2e}
\usepackage{array}
\usepackage[caption=false,font=normalsize,labelfont=sf,textfont=sf]{subfig}
\usepackage{textcomp}
\usepackage{stfloats}
\usepackage{url}
\usepackage{verbatim}
\usepackage{graphicx}
\usepackage{cite}
\usepackage{caption}
\usepackage{float} 
\usepackage{tabularx}
\usepackage{stackengine}
\usepackage{multirow}
\usepackage{multicol}
\usepackage[export]{adjustbox}
\usepackage[normalem]{ulem}
\usepackage{multirow}
\usepackage{makecell}
\usepackage{booktabs}  % for \Xhline
\useunder{\uline}{\ul}{}
\begin{document}

\title{SANR: Scene-Aware Neural Representation for Light Field Image Compression with Rate-Distortion Optimization}

\author{Gai Zhang,~\IEEEmembership{Student Member,~IEEE}, Xinfeng Zhang*,~\IEEEmembership{Senior Member,~IEEE}, Lv Tang,~\IEEEmembership{Student Member,~IEEE}, Hongyu An, Li Zhang,~\IEEEmembership{Senior Member,~IEEE}, Qingming Huang,~\IEEEmembership{Fellow,~IEEE}
        % <-this % stops a space
\thanks{Gai Zhang, Xinfeng Zhang, Lv Tang, Hongyu An, and Qingming Huang are with the School of Computer Science and Technology, University of Chinese Academy of Sciences, Beijing, China (Email: zhanggai16@mails.ucas.ac.cn, xfzhang@ucas.ac.cn, luckybird1994@gmail.com, anhongyu22@mails.ucas.ac.cn, qmhuang@ucas.ac.cn). Xinfeng Zhang is the corresponding author. 

Li Zhang is with the Advanced Video Group (AVG), Bytedance Inc.,San Diego, CA 92122 USA (Email:lizhang.idm@bytedance.com).}% <-this % stops a space
%\thanks{Manuscript received April 19, 2021; revised August 16, 2021.}}
}

\maketitle

\begin{abstract}
Light field images capture multi-view scene information and play a crucial role in 3D scene reconstruction. However, their high-dimensional nature results in enormous data volumes, posing a significant challenge for efficient compression in practical storage and transmission scenarios. Although neural representation-based methods have shown promise in light field image compression, most approaches rely on direct coordinate-to-pixel mapping through implicit neural representation (INR), often neglecting the explicit modeling of scene structure. Moreover, they typically lack end-to-end rate-distortion optimization, limiting their compression efficiency. To address these limitations, we propose SANR, a Scene-Aware Neural Representation framework for light field image compression with end-to-end rate-distortion optimization. For scene awareness, SANR introduces a hierarchical scene modeling block that leverages multi-scale latent codes to capture intrinsic scene structures, thereby reducing the information gap between INR input coordinates and the target light field image. From a compression perspective, SANR is the first to incorporate entropy-constrained quantization-aware training (QAT) into neural representation-based light field image compression, enabling end-to-end rate-distortion optimization. Extensive experiment results demonstrate that SANR significantly outperforms state-of-the-art techniques regarding rate-distortion performance with a 65.62\% BD-rate saving against HEVC.
\end{abstract}


\section{Introduction}
\IEEEPARstart{L}{ight} field images capture both spatial and angular information of a scene, forming a 4D representation that enables comprehensive scene understanding \cite{lf2016,mazhan17,levoy2023light}. They are typically acquired using lenslet-based plenoptic cameras, which sample light rays through a main lens and a microlens array. Each captured view is referred to as a sub-aperture image (SAI). Due to their rich multi-view characteristic, light field images are essential for applications such as 3D scene reconstruction \cite{jaykuo, cai2018ray}, depth estimation \cite{wang2015occlusion,jeon2019pami,mishiba2020fast}, and virtual reality \cite{yu2017light,overbeck2018system,zhangqi2023vr}. However, their high-dimensional nature results in massive data volumes, posing significant challenges for storage, transmission, and processing. Hence, highly efficient compression techniques are critical for the practical deployment of light field images.

Traditional image codecs such as JPEG \cite{1993JPEG} and BPG \cite{2015BPG} are ill-suited for light field data due to their inability to exploit inter-view correlations. To overcome this limitation, prior work \cite{monteiro2016light} organized SAIs into pseudo-video sequences and applied video codecs such as HEVC \cite{hevc}, yet these methods often fail to fully leverage the underlying geometric structure of light fields. Subsequent approaches \cite{astola2018wasp,alves2020jpeg} improved prediction accuracy by incorporating disparity maps and sparse view synthesis. Recently, implicit neural representation (INR) has emerged as a powerful paradigm for compact scene modeling, with NeRF \cite{mildenhall2020nerf} and its extensions applied to images \cite{dupont2021coin}, videos \cite{chen2021nerv}, and light fields \cite{shi2022distilled}. INR-based light field compression methods \cite{wang2022light,jiang2022untrained,shi2023learning} achieve high rate-distortion performance through continuous coordinate-to-signal mapping. However, existing approaches typically perform this mapping without explicitly capturing scene priors, resulting in a significant information gap between input coordinates and the target output. Furthermore, the reconstruction and compression modules are often decoupled, preventing end-to-end rate-distortion optimization and limiting overall efficiency.

To address these limitations, we propose SANR: scene-aware neural representation for light field image compression with rate-distortion optimization. SANR leverages scene information and end-to-end rate-distortion optimization to improve the compression performance. First, SANR introduces a scene-aware neural representation that explicitly considers the scene's content to reduce the information gap between the input coordinates of INR and the target image. SANR adopts a hierarchical scene modeling block to achieve this, which can capture multi-frequency scene details by utilizing latent codes at different scales. Second, to enable the end-to-end rate-distortion optimization, SANR employs an entropy-constrained quantization aware training scheme that takes advantage of the estimated entropy of the latent codes and parameters of the neural network. This scheme optimizes the rate-distortion performance by incorporating entropy constraints during training. Extensive experiments on benchmark datasets demonstrate that SANR achieves superior compression efficiency and visual quality compared with state-of-the-art methods.

The main contributions of this paper are summarized as follows:
\begin{itemize}
    \item We propose SANR, a scene-aware neural representation framework for light field image compression, which explicitly incorporates scene priors into the implicit neural representation to bridge the information gap between input coordinates and target light field signals, enabling more accurate reconstruction.

    \item We formulate an end-to-end rate-distortion optimization scheme through entropy-constrained quantization-aware training, where both the parameters of implicit neural networks and the latent codes are jointly optimized to minimize the rate-distortion objective loss.

    \item We conduct comprehensive experiments on benchmark datasets (EPFL and HCI), demonstrating that SANR achieves state-of-the-art performance with significant 65.62\% BD-rate reduction against HEVC, validating its effectiveness in light field image compression.
\end{itemize}

\section{RELATED WORK}
We briefly review existing research on light field image compression, categorized into conventional, transform-based, and learning-based approaches, followed by a discussion on INR and their emerging role in light field image compression.

\subsection{Existing Light Field Image Compression}
Early light field compression methods adapted video codecs by arranging SAIs into pseudo-video sequences \cite{monteiro2016light,dai2015lenselet,li2017pseudo}. While effective in exploiting inter-view redundancy, these methods neglect the geometric structure of light fields, treating them as generic video content and thus limiting compression efficiency.

To better exploit spatial-angular correlations, transform-based methods leverage geometric priors for more structured representation. Examples include depth warping \cite{astola2018wasp}, 4D-DCT in JPEG Pleno \cite{alves2020jpeg}, shearlet \cite{ahmad2020shearlet}, graph transform \cite{rizkallah2019geometry}, and graph-guided optimization \cite{hou2023TMM}. Despite improved performance, these methods rely on handcrafted transforms and are sensitive to inaccuracies in disparity estimation or graph construction.

More recently, deep learning has enabled end-to-end optimization of light field compression. Some approaches focus on view synthesis from sparse inputs \cite{liu2022multi,zhang2022light}, while others use disparity-guided prediction \cite{hu2021multiple,huang2023prediction} or feature decoupling \cite{tong2022sadn}. However, these methods often struggle to model the full 4D structure due to the high dimensionality and complex correlations.

\subsection{Implicit Neural Representation}
Implicit Neural Representations (INRs) have emerged as a powerful alternative to discrete signal representation, parameterizing continuous signals (images, videos, 3D scenes) as coordinate-to-value mappings via neural networks \cite{stanley2007compositional,sitzmann2020implicit}. A key advantage of INRs is their ability to implicitly encode high-frequency details and structural priors, enabling high-fidelity reconstruction from compact network weights.

In 2D image compression, COIN \cite{dupont2021coin} pioneers the use of INRs by fitting images with MLPs and compressing the weights. Subsequent works improve representation capacity through better architectures \cite{chen2021learning, zhang2022implicit} and quantization strategies \cite{gordon2023quantizing,strumpler2022implicit}. For video, NeRV \cite{chen2021nerv} adopts a lightweight CNN to parameterize frames along time, achieving efficient compression via model size reduction. Variants enhance performance with hierarchical structures \cite{li2022nerv}, non-linear temporal modeling \cite{chen2023hnerv}, and hybrid network designs \cite{kwan2023hinerv}.

In 3D vision, Neural Radiance Fields (NeRF) \cite{mildenhall2020nerf} revolutionize novel view synthesis by representing scenes as continuous radiance fields. Extensions such as Mip-NeRF \cite{barron2021mip} address aliasing, while Instant NGP \cite{muller2022instant} accelerates training via hash encoding. These advances highlight the potential of INRs for compact, continuous scene modeling.

Inspired by these successes, recent works explore INRs for light field image compression. Feng {\em et al.} \cite{feng2021signet,feng2022neural} use INRs to reconstruct light fields from sparse views, demonstrating their ability to model angular continuity. Wang {\em et al.} \cite{wang2022light} propose an MLP-NeRV hybrid to represent SAIs and apply model compression techniques for light field image compression. Shi {\em et al.} \cite{shi2022distilled} adapt NeRF for light field synthesis and employ tensor decomposition to reduce model size. Jiang {\em et al.} \cite{jiang2022untrained} use a GRU to model view variations and a deep decoder for spatial details, while Shi {\em et al.} \cite{shi2023learning} introduce modulated CNN kernels for view-specific synthesis.

Despite these efforts, existing INR-based methods share two key limitations. First, they typically map input coordinates and noise to pixel values without explicit incorporation of scene structure, leading to a significant information gap between the network input and the target signal. This limits modeling accuracy, especially for complex scenes with occlusions or texture variations. Second, most methods adopt a two-stage pipeline: first training the INR, then compressing its weights. They lack end-to-end rate-distortion optimization, failing to jointly optimize representation fidelity and compression bitrate.

In contrast, our SANR framework bridges this information gap by explicitly integrating scene priors into the INR through hierarchical scene modeling and formulating a unified rate-distortion objective with quantization-aware training. This enables joint optimization of scene-aware representation and compression efficiency, leading to superior performance.

\begin{figure*}[htbp]
    \centering
    \includegraphics[width=\textwidth]{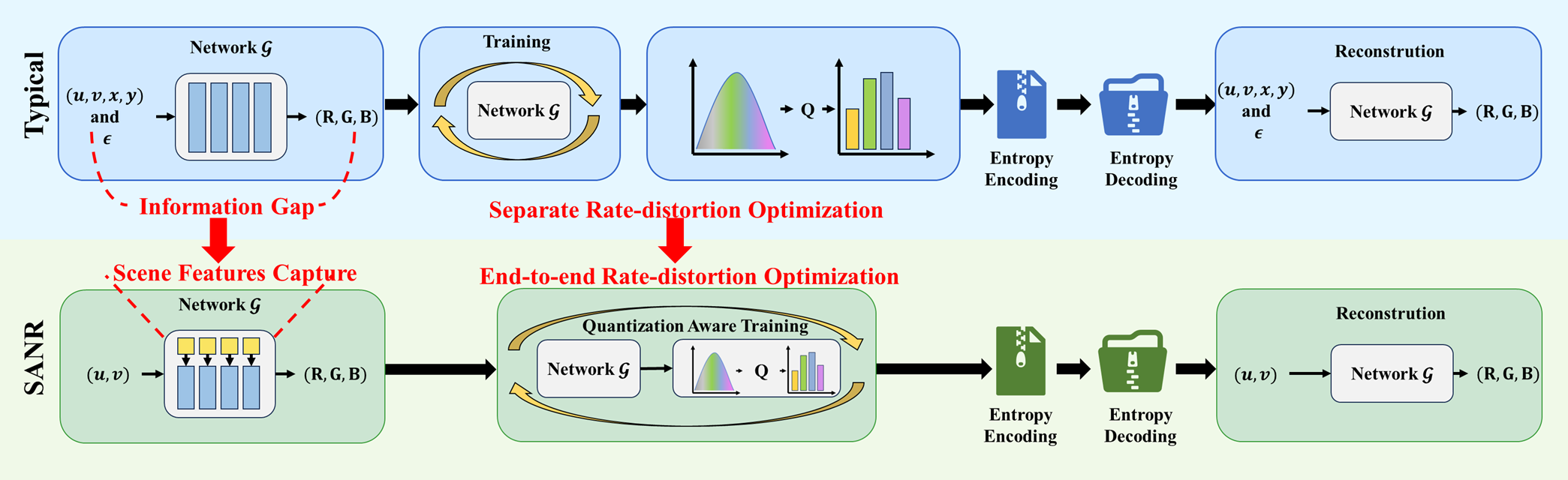}
    \caption{This figure shows the typical INR-based and the SANR light field image compression pipeline. For the typical method, $\mathcal{G}$ is the neural network for mapping pixel coordinates $(u,v,x,y)$ and given noise $\epsilon$ to the corresponding $(R,G,B)$ values. SANR takes the input of angle coordinate $(u, v)$ and realises the image-level mapping. The typical method omits the scene information capture and cannot reduce the information gap between the input and image signals. SANR gives the network $\mathcal{G}$ with the ability to capture scene features. During the encoding stage, the typical method first trains the network $\mathcal{G}$ to fit the target light field image and then quantizes the trained network $\mathcal{G}$. The separate rate-distortion optimization leads to suboptimal compression performance. SANR realizes the end-to-end rate-distortion optimization with the QAT and Stochastic Gumbel Annealing (SGA) \cite{yang2020improving}. For decoding, both methods need to entropy decode the trained parameters of the network $\mathcal{G}$ and then execute forward propagation to reconstruct light field images.}
    \label{fig:pipeline}
\end{figure*}

\section{Proposed Method}

\subsection{Preliminaries}
Typical INR-based light field image compression follows a pipeline of network training, quantization, and entropy coding, as illustrated in Fig.~\ref{fig:pipeline}. The core idea is to use a neural network $\mathcal{G}$, typically composed of MLPs or CNNs, to map input coordinates and a fixed noise $\epsilon$ to corresponding signal values, expressed as $\mathcal{G}(input) \rightarrow output$. Specifically, for light field images, $\mathcal{G}$ maps the 4D pixel coordinates $(u,v,x,y)$ and noise $\epsilon$ to the RGB values $(R, G, B)$. To represent the target light field, the network $\mathcal{G}$ is trained to fit the data by minimizing a reconstruction loss such as mean squared error (MSE) using gradient descent. After training, the network parameters are quantized and are further compressed through entropy coding, such as arithmetic coding \cite{witten1987arithmetic}, into a compact bitstream for storage or transmission.

At the decoding stage, the bitstream is first entropy decoded to recover the quantized network parameters. These parameters are then loaded into the network $\mathcal{G}$, which is executed with all required pixel coordinates and the same noise $\epsilon$, generated from a known seed, to reconstruct the light field image via forward propagation.

This generative approach leverages neural networks to implicitly represent the light field. However, the inputs, which consist of coordinates and random noise, do not contain meaningful image content. As a result, a significant information gap exists between the input and the target signal, as shown in Fig.~\ref{fig:pipeline}. This limits the network's ability to efficiently capture scene structure and degrades compression performance. To address this issue, SANR introduces a hierarchical scene modeling block that enables the network to explicitly capture scene features within the INR framework, thereby reducing the information gap. The design of this block is detailed in Section~\ref{HSMB}.

In the conventional INR encoding scheme, network training and compression are performed in separate stages, which prevents end-to-end rate-distortion optimization. In contrast, as shown in Fig.~\ref{fig:pipeline}, SANR integrates rate-distortion optimization into the training process by introducing Quantization-Aware Training (QAT) and incorporating the entropy of the model parameters into the loss function. After QAT, the Stochastic Gumbel Annealing (SGA) algorithm \cite{yang2020improving} is applied to further refine the rate-distortion trade-off. These components enable end-to-end optimization of both representation fidelity and compression efficiency. The implementation details are elaborated in Section~\ref{QAT}.

\begin{figure*}[htbp]
    \centering
    \includegraphics[width=\textwidth]{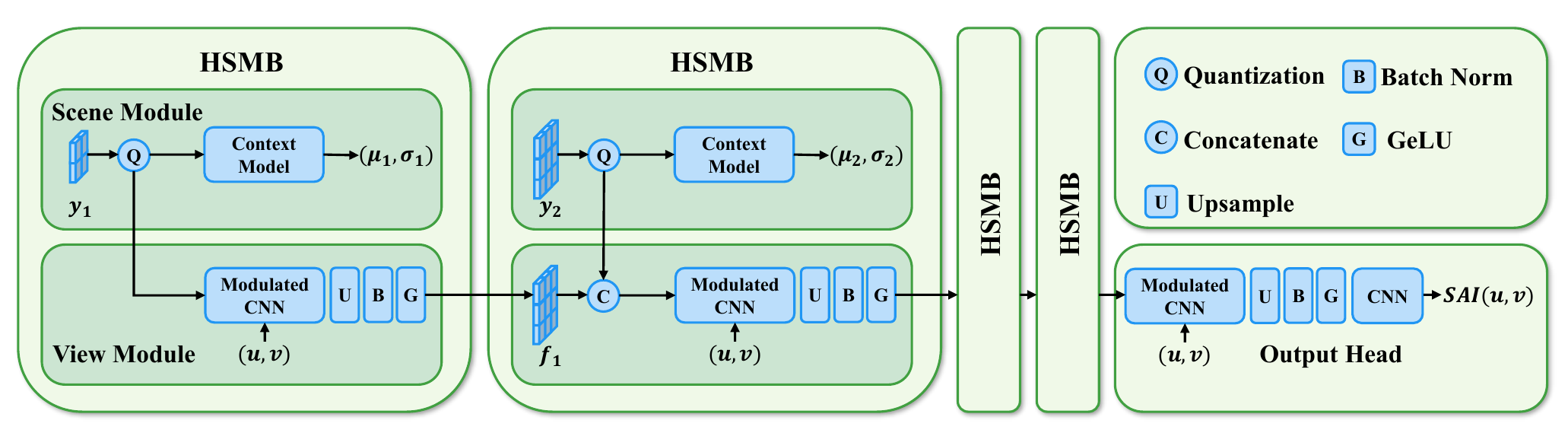}
    \caption{Architecture overview of the proposed SANR framework. The overall network structure consists of four hierarchical scene modeling blocks (HSMBs), forming the backbone of SANR. Each HSMB contains a scene module and a view module. The scene module encodes spatial-structural priors into a latent scene code $y_i$, which is modeled by a context model that estimates its distribution parameters (mean $\mu_i$ and standard deviation $\sigma_i$) for entropy coding. The view module transforms the intermediate feature $f_i$ into the output feature using a modulated CNN, which is conditioned on the input angular coordinate $(u, v)$ to synthesize the corresponding sub-aperture image (SAI).}
\label{fig:framework}
\end{figure*}

\subsection{Hierarchical Scene Modeling Block} \label{HSMB}
The complete architecture of the proposed SANR framework is illustrated in Fig.~\ref{fig:framework}. SANR's network adjusts the network parameters based on different angle coordinates $(u, v)$ to achieve mapping of all SAIs. As shown in Fig.~\ref{fig:framework}, hierarchical scene modeling block (HSMB) serves as the backbone of SANR, with four such blocks stacked hierarchically to progressively capture and refine scene features. Each HSMB comprises two key components: a \textit{scene module} and a \textit{view module}. The scene module encodes spatial-structural priors into latent scene codes $y_i$, which are learned during training and entropy-coded for compression. The context model associated with each $y_i$ estimates its distribution parameters, mean $\mu_i$ and standard deviation $\sigma_i$, based on previously decoded neighbors, facilitating efficient entropy coding. The view module, built upon a modulated CNN, transforms intermediate features into output features conditioned on the angular coordinate $(u, v)$, enabling the synthesis of any SAI on demand. This design decouples scene structure from view synthesis, allowing the network to learn rich spatial representations while maintaining view adaptability. The following sections detail the structure of each module.

\begin{figure}
    \centering
    \resizebox{\columnwidth}{!}{
        \includegraphics{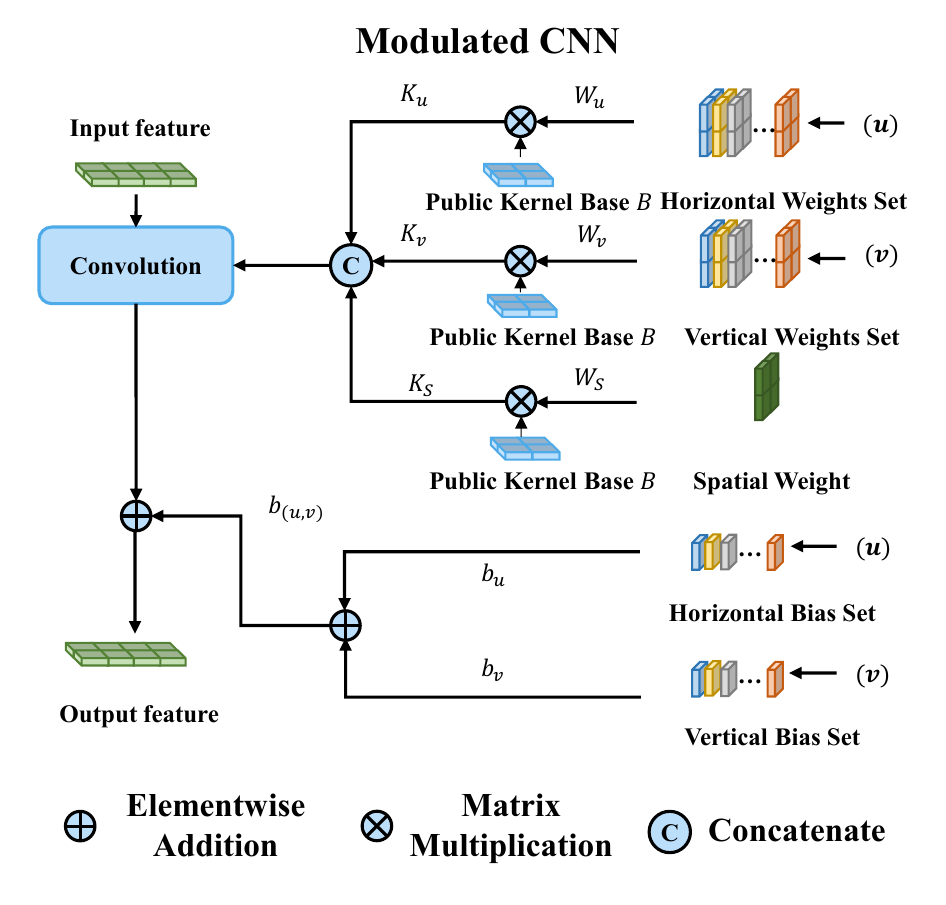}
    }
    \caption{The architecture of the modulated CNN. The angular coordinate $(u, v)$ is used to generate modulation parameters: horizontal and vertical weights $W_u$, $W_v$ and biases $b_u$, $b_v$. These, together with spatial weights $W_s$, are applied through tensor outer products with a shared kernel base $B$. The resulting kernels are concatenated and used in convolution with the input feature. The output is obtained by adding the modulated bias $b_{u,v}$, allowing the network to dynamically generate SAI views on demand.}
    \label{fig:modulated_cnn}
\end{figure}

\subsubsection{View Module}
The view module is responsible for generating view-specific features from shared latent scene codes. As depicted in Fig.~\ref{fig:framework}, it processes the intermediate feature $f_i$ through a modulated CNN, followed by upsampling, batch normalization, and GeLU activation to produce the output feature. The modulated CNN is the core component that enables conditional feature generation based on the input angular coordinate $(u, v)$.

As shown in Fig.~\ref{fig:modulated_cnn}, the modulated CNN dynamically generates its convolution kernel and bias based on $(u, v)$. The resulting kernel consists of three components: the spatial kernel $K_S$, the horizontal kernel $K_u$, and the vertical kernel $K_v$. The spatial kernel $K_S$ serves as the base component, capturing common spatial patterns across all views and providing the fundamental modeling capacity for the entire light field. To incorporate angular variation, $K_u$ and $K_v$ are selected from horizontal and vertical kernel sets according to the input $(u, v)$. These three kernels are concatenated along the output channel dimension to form the complete modulated convolution kernel. Similarly, the bias is modulated by selecting horizontal and vertical bias components $b_u$ and $b_v$ from corresponding bias sets, further enriching view-specific rendering information.

The view module utilizes the tensor decomposition technique for kernel compression in the kernel sets. Specifically, as illustrated in Fig.~\ref{fig:modulated_cnn}, the spatial kernel $K_S$ is decomposed into a shared kernel base $B$ and a coefficient tensor $W_S$ via tensor outer product:
\begin{equation}
    K_S = W_S \otimes B,
\end{equation}
where $K_S$ is of size $C^{S}_{out} \times C_{in} \times k \times k$, $C_{in}$ is the input channel number, $C^{S}_{out}$ is the output channel number, $B$ is of size $r \times k \times k$, and $W_S$ is the coefficient of size $C^{S}_{out} \times C_{in} \times r$. The symbol $\otimes$ denotes the tensor outer product, and $r$ is the number of bases in $B$. $K_u$ and $K_v$ are similar to $K_S$, but only one different operation for choosing the weights $W_u$ and $W_v$. $W_u$ and $W_v$ are chosen from the view weights set by the input angular coordinate $(u, v)$.

\subsubsection{Scene Module}

\begin{figure}
    \centering
    \resizebox{\columnwidth}{!}{
        \includegraphics{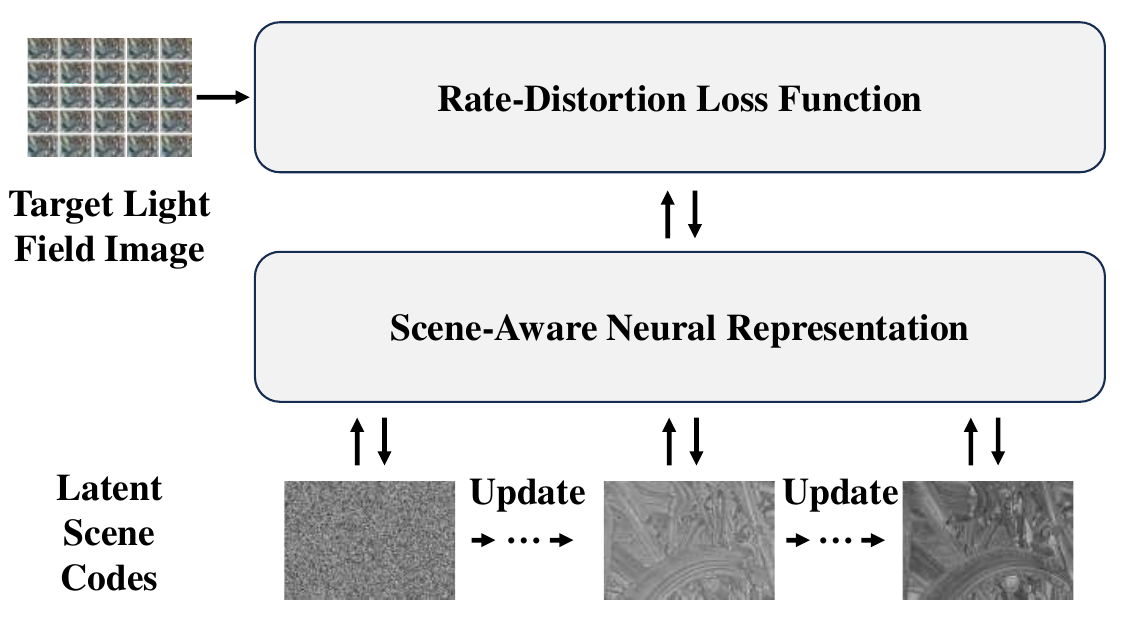}
    }
    \caption{The figure illustrates how latent scene codes progressively capture scene features during training.}
    \label{fig:scene code}
\end{figure}

To compensate for the information gap between the input angle coordinates $(u, v)$ and the corresponding reconstructed images, we introduce the scene module in HSMB, as depicted in Fig.\ref{fig:framework}. The core idea is to introduce latent scene codes to learn the spatial features of the scene for accurate reconstruction of details. The latent scene code is denoted as $y_i$, where $i \in \{ 1, 2, 3, 4\}$ represents the layer index of HSMB. The latent scene codes are initially assigned random values, similar to network parameters. During training, these codes are progressively updated through backpropagation and gradient descent, resulting in a continuous reduction of the loss function. The latent scene codes gradually learn the scene features in the target light field image as shown in Fig.\ref{fig:scene code}. These latent codes have dimensions of $1 \times C_l \times h_i \times w_i$. $C_l$ is the manually set number of latent scene code channels. The height $h_i$ and width $w_i$ of the latent codes are computed based on the original height $h$ and width $w$ of the input SAI using the following formulas:
\begin{equation}
h_i = \lfloor h / 2^{5-i} \rceil, \quad w_i = \lfloor w / 2^{5-i} \rceil.
\end{equation}
During training, the latent code $y_i$ is quantized by adding uniform noise $n \in (-0.5, 0.5)$, allowing for gradient-based optimization. As shown in Fig.\ref{fig:framework}, the feature $f_{i-1}$ will be concatenated with the quantized latent code $y_i$, and the fused feature will be sent to the view module for the next feature $f_i$ generation.

\begin{figure}
    \centering
\begin{tabular}{@{} c @{}}

  \begin{tabular}{@{} c c c c @{}}
    \textbf{Ground} & \textbf{w/o large} & \textbf{w/o small} & \textbf{w/}  \\
    \textbf{Truth} & \textbf{scale latent} & \textbf{scale latent} & \textbf{all latents} \\
    \includegraphics[width=0.1\textwidth]{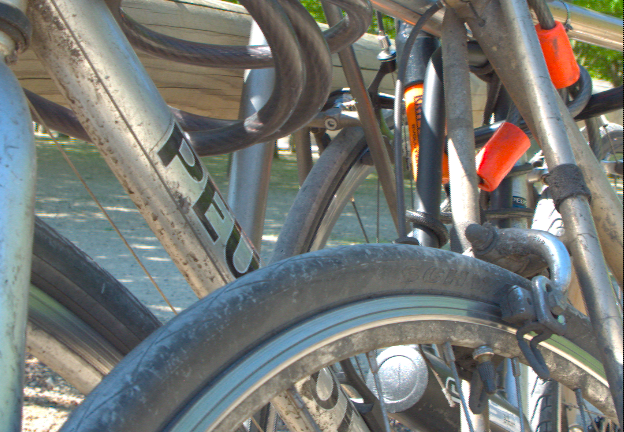} &
    \includegraphics[width=0.1\textwidth]{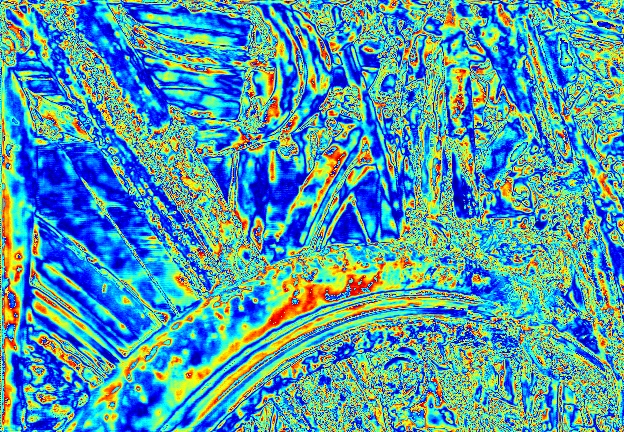} &
    \includegraphics[width=0.1\textwidth]{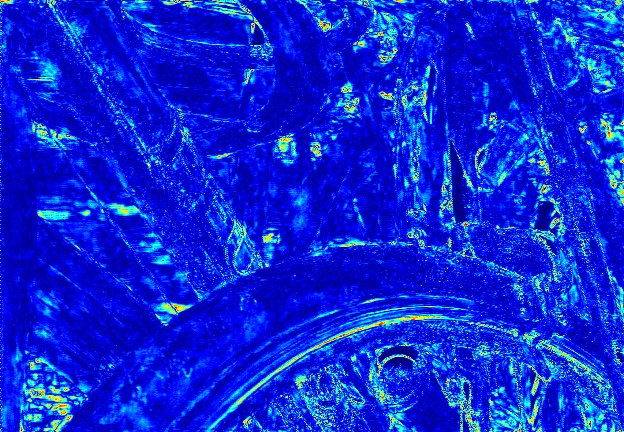} &
    \includegraphics[width=0.1\textwidth]{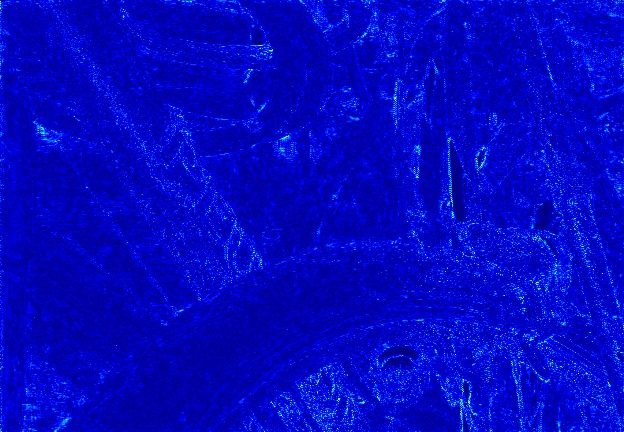} \\

    \includegraphics[width=0.1\textwidth]{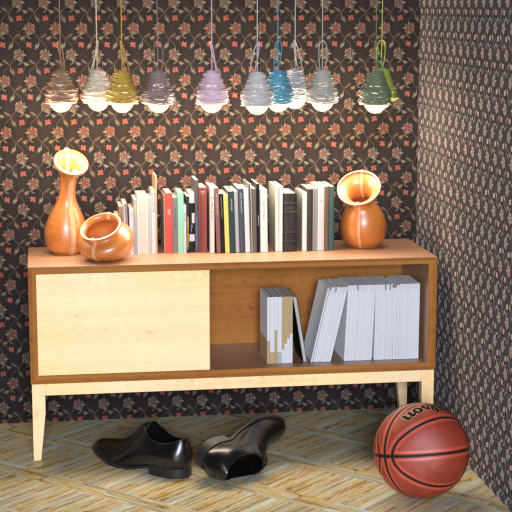} &
    \includegraphics[width=0.1\textwidth]{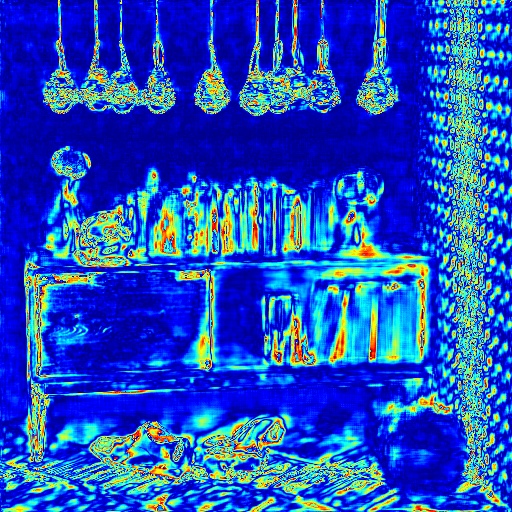} &
    \includegraphics[width=0.1\textwidth]{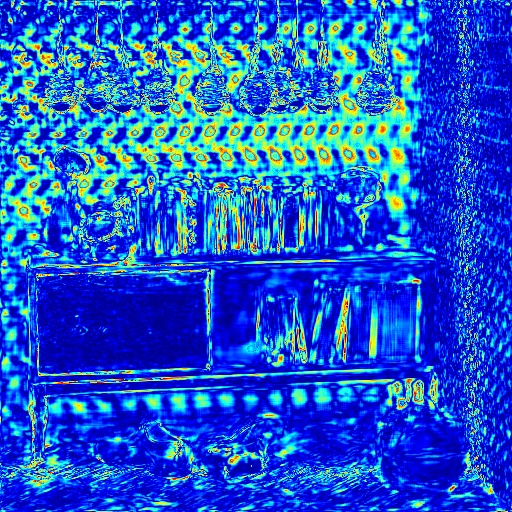} &
    \includegraphics[width=0.1\textwidth]{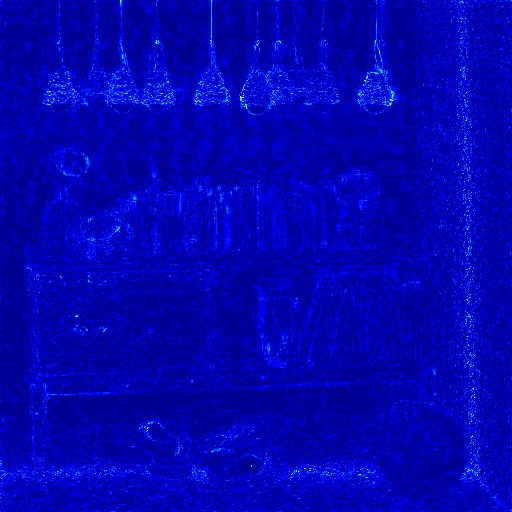}
  \end{tabular}
\end{tabular}

    \caption{The average error maps for SANR without large-scale latent codes $y_3, y_4$ and without small-scale latent codes $y_1, y_2$.}
    \label{fig:scene visual}
\end{figure}

Furthermore, by utilizing latent codes with different spatial sizes, we aim to capture information across various frequency bands, enabling effective representation of scene details at multiple scales. This multi-scale approach allows for accurate modeling of a scene’s complex and diverse structures. To provide visual evidence of SANR’s effectiveness in capturing scene features, we present the results in Fig.~\ref{fig:scene visual}. In this figure, we show the visualizations of the average error maps from the trained network when using different subsets of latent codes, evaluated on the \textit{Bikes} sequence from the EPFL dataset and the \textit{sideboard} sequence from the HCI dataset. Upon examining the visualizations, it becomes evident that the latent codes successfully capture essential scene features. Notably, we observe that images reconstructed by networks lacking large-scale latent codes $y_3, y_4$ exhibit larger errors in the high-frequency components of the foreground, while those reconstructed without small-scale latent codes $y_1, y_2$ show increased errors in the background. These results demonstrate that the multi-scale latent codes provide a comprehensive and informative encoding of the scene, ensuring that critical details are preserved across different spatial scales. By effectively capturing scene features, the hierarchical scene modeling block enhances the fidelity of the reconstructed SAIs.

\begin{figure}
    \centering
    \resizebox{\columnwidth}{!}{
        \includegraphics{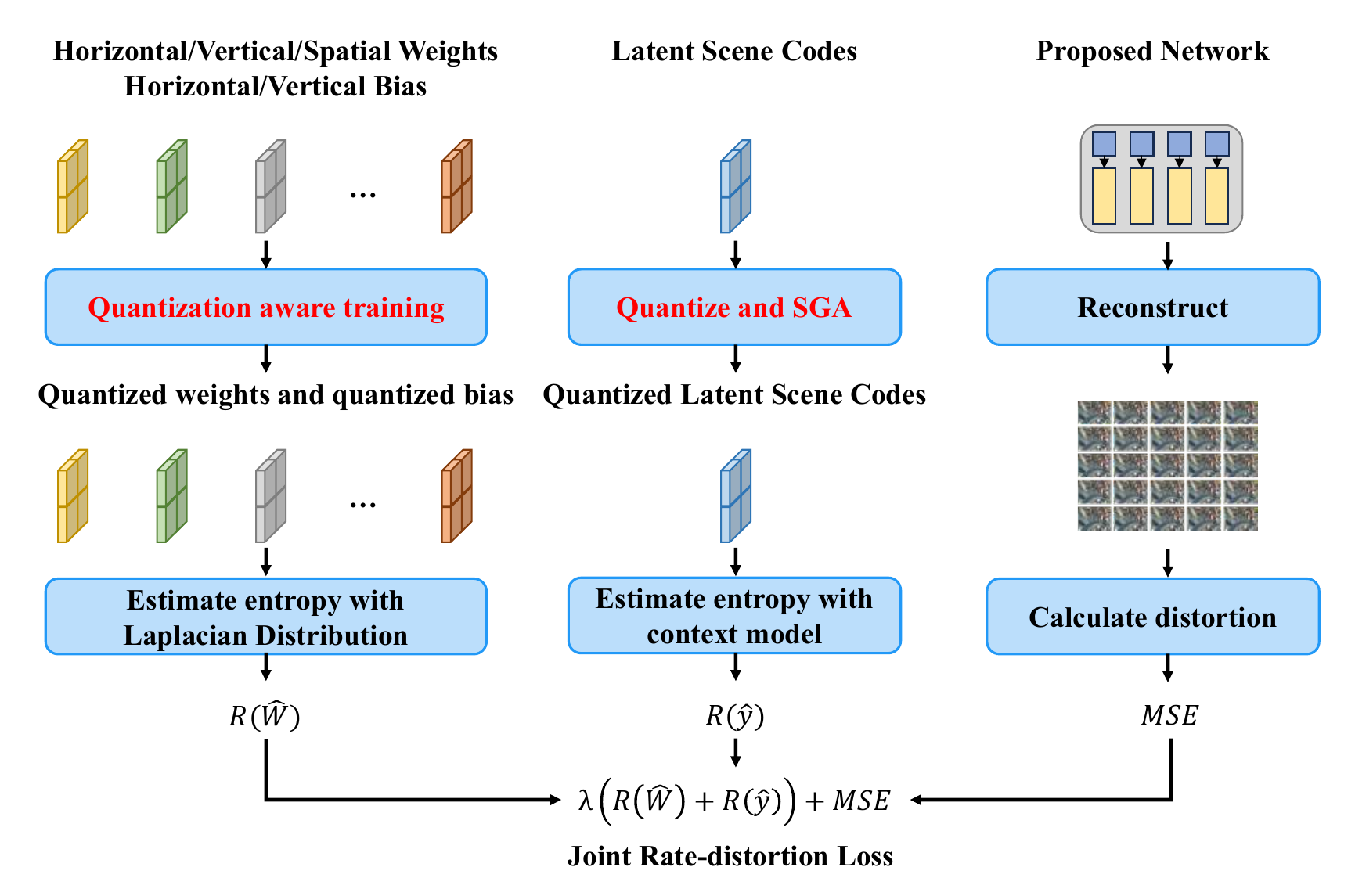}
    }
    \caption{With the QAT technique, we can quantize the network parameter weights and estimate their entropy $R(\hat{W})$ during training. Similarly, we quantize the latent scene codes by adding uniform noise and applying SGA. We estimate the entropy $R(\hat{y})$ of the latent scene codes with the context model. We utilize mean-square-error (MSE) to evaluate the distortion between reconstructed light field images and target light field images. Finally, we get the end-to-end rate-distortion loss function and realize end-to-end rate-distortion optimization with it.}
    \label{fig:loss}
\end{figure}

% \vspace{-0.5cm}

\subsection{End-to-End Rate Distortion Optimization}\label{QAT}
As depicted in Fig.\ref{fig:pipeline}, typical INR-based light field image compression techniques leverage network compression methods to reduce the redundancy of the fitted network. During the compression process, the network parameters are quantized and subsequently entropy-coded using established methods such as arithmetic coding. This compression approach effectively reduces the data size while preserving an acceptable level of reconstruction performance. However, post-training quantization (PTQ) only allows for unilateral adjustment of the model size without optimizing reconstruction performance. Consequently, PTQ fails to achieve optimal rate-distortion trade-offs, leading to a key challenge: how to jointly optimize reconstruction fidelity and model quantization for improved rate-distortion performance.

To address this challenge, we realize an end-to-end rate-distortion optimization. As illustrated in Fig.~\ref{fig:loss}, we adopt quantization-aware training (QAT) to compress network parameters, enabling entropy estimation $R(\hat{W})$ during optimization. Additionally, we apply uniform noise injection and SGA to quantize the latent scene codes during training, and estimate their entropy $R(\hat{y})$ using a context model. The MSE distortion between the reconstructed and original light field images is computed. By combining rate and distortion terms, we formulate a rate-distortion loss that enables end-to-end optimization during training. The specific entropy estimation methods are detailed below.

\begin{figure}[htbp]
    \centering
\begin{tabular}{@{} c @{}}

  \begin{tabular}{@{} c @{} c @{} }

    \includegraphics[width=0.25\textwidth]{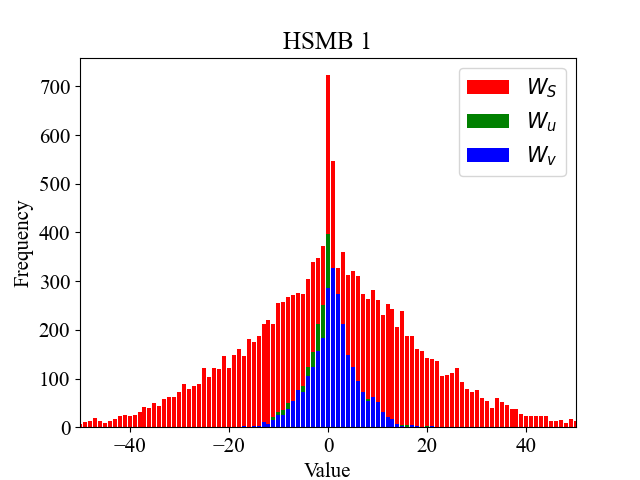} &
    \includegraphics[width=0.25\textwidth]{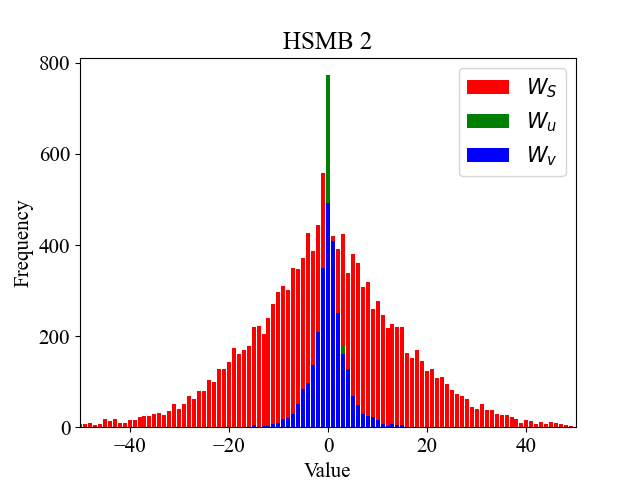} \\
    \includegraphics[width=0.25\textwidth]{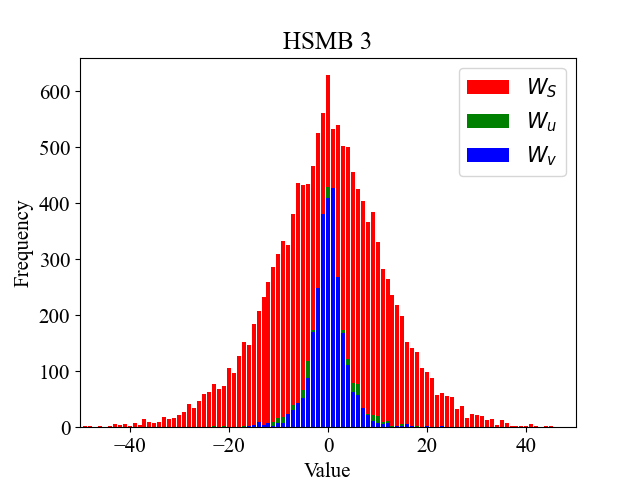} &
    \includegraphics[width=0.25\textwidth]{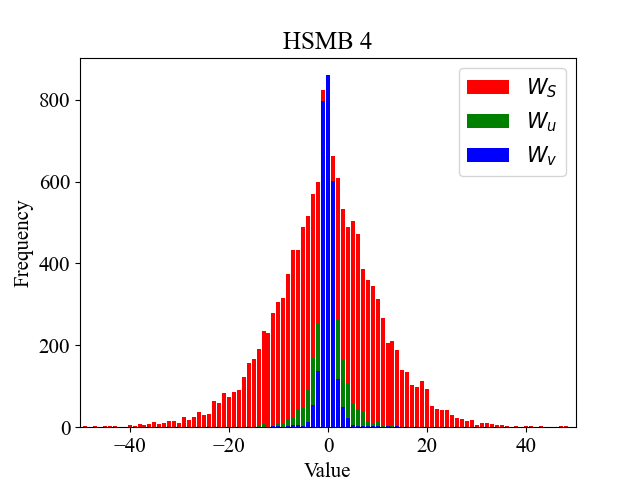}
    \\

  \end{tabular}
\end{tabular}

\caption{Histogram of the quantized weights from a trained model for \textit{Bikes} sequence. Spatial weights \textbf{$W_S$}, horizontal weights \textbf{$W_u$}, and vertical weights \textbf{$W_v$} all have a Laplacian distribution.}
    \label{fig:hist}
\end{figure}

% \vspace{-0.5cm}

\subsubsection{Network Parameter Weights}
We apply the QAT technique to network parameter weights for rate-distortion optimization. QAT is a model compression approach that simultaneously quantizes and optimizes the model's performance. QAT quantifies the model parameters while updating them with gradient descent and backpropagation algorithms. However, directly rounding the parameters during training may cause the gradients to zero throughout the network, which hampers effective training. To solve this issue, we employ the straight-through estimator (STE) \cite{ste} technique to quantize the parameters during training.

We focus on quantizing the weights of view module components, specifically the spatial kernel weights $W_S$, horizontal kernel weights $W_u$, vertical kernel weights $W_v$, and modulated bias $b_{u}$, $b_{v}$. For other components that include kernel bases $B$, context model, and the last decoding layer, we extend QAT to these components. But QAT on them significantly degrades distortion performance and consequently degrades rate-distortion performance. Because these components comprise less than 2\% of total parameters, we apply 16-bit uniform quantization for them after training, and we directly save these parameters as 16 bits in the bitstream.

We first utilize QAT to train a model with the following loss:
\begin{equation}
MSE = \sum_{u,v}(x_{(u,v)}-\hat{x}_{(u,v)})^{2},
\end{equation}
where $x_{(u,v)}$ represents the original SAI at the angular position $(u,v)$, and $\hat{x}_{(u,v)}$ is the corresponding reconstructed SAI. We analyze the data distribution of the quantified network parameters. Fig.\ref{fig:hist} illustrates histograms of the quantized weights obtained from the QAT model applied to the \textit{Bikes} sequence. The histograms demonstrate that the distribution of the weight data follows a Laplacian distribution pattern. Based on this observation, we estimate the entropy of the quantized weight data using the Laplacian distribution:

\begin{equation}
R(\hat{w}) = -\log_{2} \int_{\hat{w}-0.5}^{\hat{w}+0.5} \mathcal{L}(\mu_w, \sigma_w) dx,
\end{equation}
where $\mathcal{L}$ is the Laplace distribution probability density function, $\mu_w$ and $\sigma_w$ represent the mean and standard deviation of the corresponding weight $w$ data respectively. $\mu_w$ and $\sigma_w$ are also needed for entropy decoding. As these parameters consume negligible bits, they will be directly transmitted as 32 bits.

\subsubsection{Latent Scene Codes}
\begin{figure}
    \centering
    \resizebox{\columnwidth}{!}{
        \includegraphics{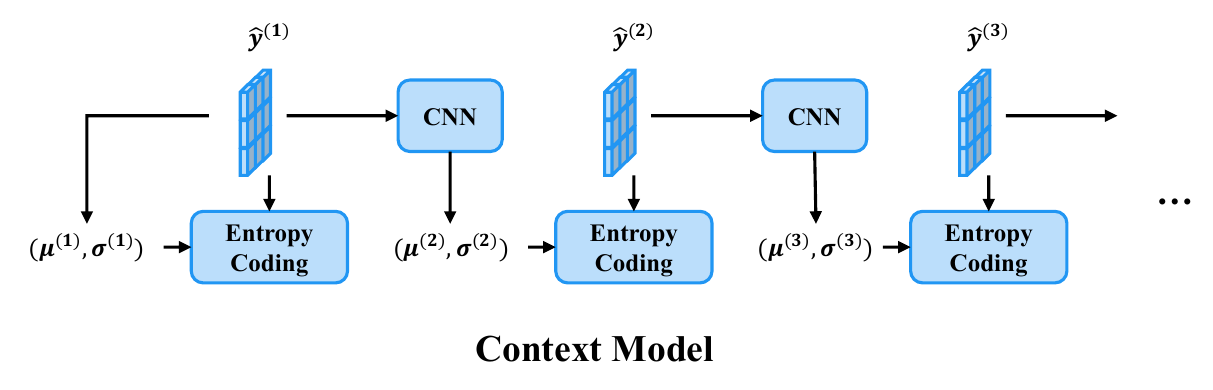}
    }
    \caption{ The context model predicts the hyperparameters $(\mu^{(c)}, \sigma^{(c)})$ of the latent code $\hat{y}^{(c)}$ based on previously decoded channel $\hat{y}^{(c-1)}$, enabling efficient entropy coding.}
    \label{fig:context_model}
\end{figure}

The latent codes $y$ also need to be encoded. To compress the latent codes $y$ efficiently, we adopt a channel-wise autoregressive context model that captures dependencies across latent code channels, as illustrated in Fig.\ref{fig:context_model}. Unlike conventional element-wise context models that predict distribution parameters sequentially for each element, our approach processes latent code channels in a sequential order, where the predicted distribution parameters for each channel is conditioned on previously decoded channels.

Specifically, the $C$-channel latent tensor $y \in \mathbb{R}^{H \times W \times C}$ is split along the channel dimension into $C$ individual slices $\{y^{(1)}, y^{(2)}, \dots, y^{(C)}\}$. The first channel $y^{(1)}$ directly uses statistical mean $\mu^{(1)}$ and standard deviation $\sigma^{(1)}$ for entropy coding. For each subsequent channel $y^{(c)}$ ($c = 2, \dots, C$), we utilize the decoded values of the previous channel $\hat{y}^{(c-1)}$ as contextual information to conditionally predict the distribution parameters. A lightweight context model $Ctx$, implemented as a stack of three $3\times3$ convolutional layers with ReLU activations, is employed to transform the previous channel's representation into the mean $\mu^{(c)}$ and standard deviation $\sigma^{(c)}$ of a conditional Laplace distribution for the current channel.

Herein, the total rate cost $R(\hat{y})$ of the quantized latent codes $\hat{y}$ can be computed as:
\begin{equation}
\begin{aligned}
    R(\hat{y}) &= -\sum_{c=1}^{C} \sum_{i \in \Omega^{(c)}} \log_2 q(\hat{y}_i^{(c)} \mid \hat{y}^{(c-1)}) \\
               &= -\sum_{c=1}^{C} \sum_{i \in \Omega^{(c)}} \log_2 \int_{\hat{y}_i^{(c)} - 0.5}^{\hat{y}_i^{(c)} + 0.5} \mathcal{L}(u; \mu_i^{(c)}, \sigma_i^{(c)})  du,
\end{aligned}
\end{equation}
where $\Omega^{(c)}$ denotes the spatial support of the $c$-th channel.

Therefore, to achieve optimal rate-distortion performance during the QAT process, the model is trained using the following loss function:
\begin{equation}
    L = \sum_{u,v}(x_{(u,v)} - \hat{x}_{(u,v)})^{2} + \lambda \left( R(\hat{y}) + R(\hat{w}) \right),
\end{equation}
where $\lambda$ represents the Lagrangian multiplier that balances the trade-off between distortion and total bitrate.

\section{EXPERIMENT}

\subsection{Experiment setting}

\subsubsection{Training setting}
We use PyTorch to build and train SANR. We use a dataloader that samples 500 times for each SAI from the target light field image. During training, we feed 5 SAIs into the network in each iteration and calculate the average loss. We update the rate-distortion loss every 5 iterations, and only update the distortion loss in the remaining iterations.  We choose the Adam \cite{kingma2014adam} optimizer for optimization and initialize the learning rate at 0.01. To enhance the convergence of the training process, we incorporate a learning rate schedule strategy. Specifically, if the loss function does not decrease for two consecutive epochs during evaluation, we halve the learning rate. We perform this adjustment twice and terminate the training if the learning rate is further halved. The max training epochs are 30. Then, we will take 6 epochs of fine-tuning with SGA. We utilise a single NVIDIA V100 GPU to facilitate the training process.

\subsubsection{Test datasets}
We thoroughly evaluate SANR using diverse light field images, including the widely recognized EPFL light field image dataset \cite{rerabek2016new} and the HCI light field image dataset \cite{honauer2017dataset}. 

We choose four real-world scenes from the EPFL light field dataset, including \textit{Bikes}, \textit{Danger}, \textit{FountainVincent2}, and \textit{StonePillarsOutside}. These light field images are captured using Lytro Power Tools, which decompose the images into view arrays from lenslet images. The spatial-angular resolution of the light field images is 434 × 625 × 15 × 15. We select the central 9 × 9 view array as the experimental data to eliminate vignetting effects. Furthermore, we crop the black edges on the left and top of the SAIs, resulting in a final size of 432 × 624. The EPFL light field images are known to exhibit noise, making them suitable for evaluating the robustness of the compared methods against noise artifacts.

\begin{table*}[h]
    \centering
    \renewcommand\arraystretch{1.2}
    \resizebox{\textwidth}{!}{
    \begin{tabular}{c|c|c|c|c|c|c|c|c|c}
    \hline \hline
     & \textit{Bikes} & \textit{Danger} & \textit{Fountain}  & \textit{Stone} & \textit{boxes} & \textit{sideboard} & \textit{cotton} & \textit{dino} & Average \\ \hline
     & \makecell{BD-Rate[\%]/\\BD-PSNR[dB]} 
     & \makecell{BD-Rate[\%]/\\BD-PSNR[dB]} 
     & \makecell{BD-Rate[\%]/\\BD-PSNR[dB]} 
     & \makecell{BD-Rate[\%]/\\BD-PSNR[dB]} 
     & \makecell{BD-Rate[\%]/\\BD-PSNR[dB]} 
     & \makecell{BD-Rate[\%]/\\BD-PSNR[dB]} 
     & \makecell{BD-Rate[\%]/\\BD-PSNR[dB]} 
     & \makecell{BD-Rate[\%]/\\BD-PSNR[dB]} 
     & \makecell{BD-Rate[\%]/\\BD-PSNR[dB]} \\
    \hline
    MV-HEVC & -4.07 / 0.09 & -5.83 / 0.13 & -6.51 / 0.16 & -11.20 / 0.20 & -22.74 / 0.57 & -22.34 / 0.92 & -7.39 / 0.21 & -20.71 / 0.66 & -12.59 / 0.36 \\
    VVC     & -22.77 / 0.56 & -18.86 / 0.53 & -22.33 / 0.64 & -24.95 / 0.51 & -31.21 / 0.78 & -31.90 / 1.31 & -22.56 / 0.59 & -31.66 / 1.08 & -25.77 / 0.75 \\
    JPEG-P  & -31.30 / 0.69 & -44.98 / 1.15 & -32.68 / 0.86 & -43.95 / 0.82 & 9.24 / -0.24 & 147.84 / -3.20 & 31.34 / -0.55 & 32.68 / -0.82 & 8.52 / -0.16 \\
    DCVC-DC & 32.80 / -0.55 & 61.08 / -0.92 & 46.82 / -0.79 & 118.14 / -0.96 & 0.71 / -0.01 & 13.42 / -0.40 & 29.65 / -0.56 & 30.43 / -0.76 & 41.63 / -0.61 \\
    HiNeRV  & -49.58 / 1.25 & -54.35 / 1.60 & -43.30 / 1.24 & -28.09 / 0.19 & 6.87 / -0.32 & -22.50 / 0.95 & 20.26 / -0.44 & 79.26 / -1.55 & -11.42 / 0.36 \\
    LMKNR   & -58.89 / 1.75 & -55.50 / 1.67 & -63.61 / 1.83 & -21.51 / 0.25 dB & -35.85 / 0.98 & -47.47 / 2.64 & 6.65 / 0.38 & -23.32 / 0.77 & -37.43 / 1.28 \\
    \hline
    SANR-fast    & {-67.70 / 2.15 } & {-74.92 / 2.55 } & {-68.92 / 2.29 } & {-76.58 / 2.10 } & {-54.39 / 1.33 } & {-61.52 /3.47 } & {18.32 /-0.36 } & {-27.92 /0.94 } & {-51.70 /1.81 } \\
    SANR    & \textbf{-74.85 / 2.64} & \textbf{-79.59 / 3.02 } & \textbf{-76.81 / 2.85 } & \textbf{-82.32 / 2.56 } & \textbf{-66.82 /2.04 } & \textbf{-71.41 / 4.66 } & \textbf{-29.71 / 0.52 } & \textbf{-43.44 / 1.50 } & \textbf{-65.62 / 2.47 } \\
    \hline \hline
    \end{tabular}
    }
    \caption{This table shows both BD-Rate and BD-PSNR gain of light field image compression methods against HEVC. The best BD-rate and BD-PSNR results are bolded.}
    \label{tab:bd rate}
\end{table*}

\begin{figure*}[h]
    \centering
\begin{tabular}{@{} c @{}}

  \begin{tabular}{@{} c @{} c @{} c @{} c @{} c @{} c @{}}

    \includegraphics[width=0.25\textwidth]{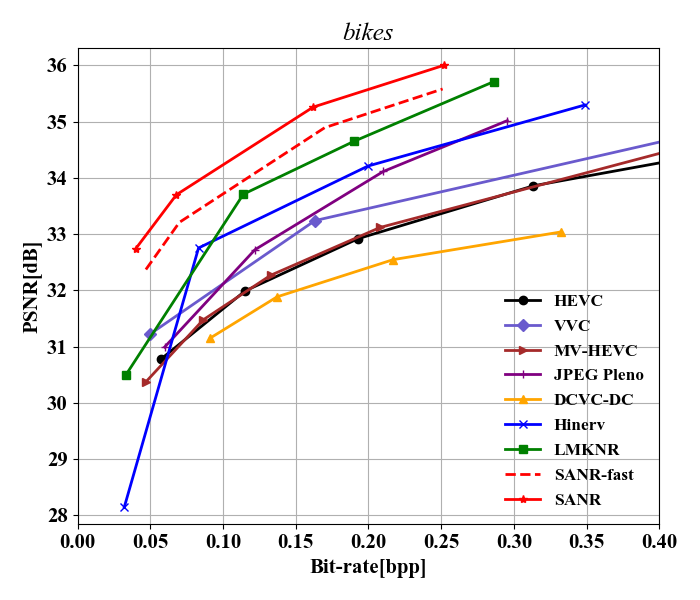} &
    \includegraphics[width=0.25\textwidth]{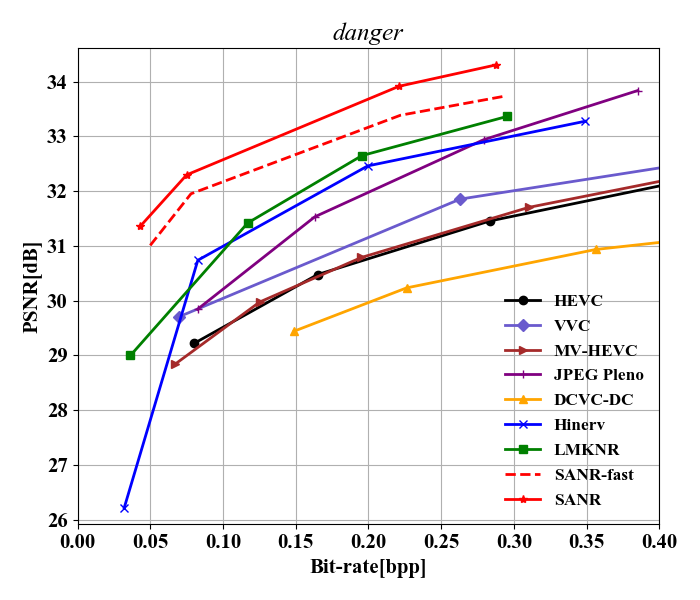} &
    \includegraphics[width=0.25\textwidth]{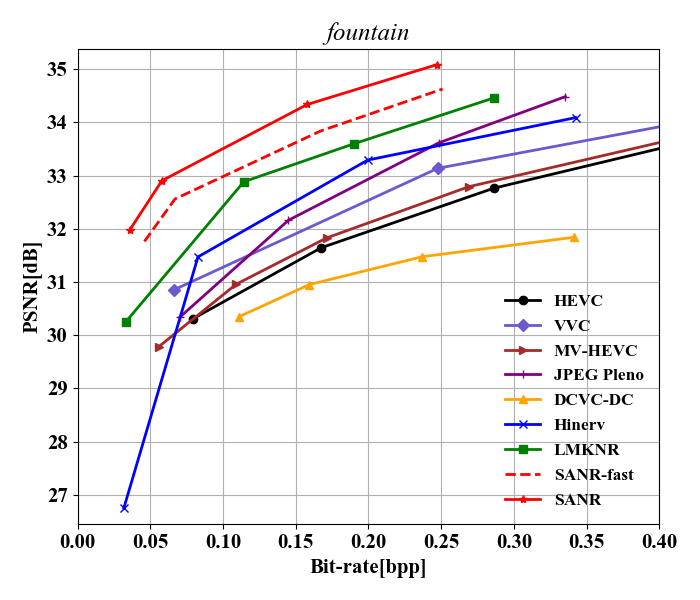} &
    \includegraphics[width=0.25\textwidth]{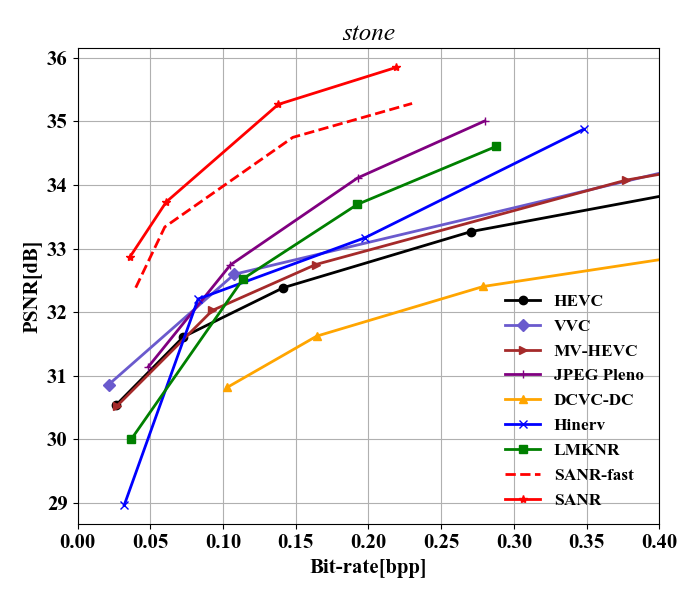}
    \\
    
    \includegraphics[width=0.25\textwidth]{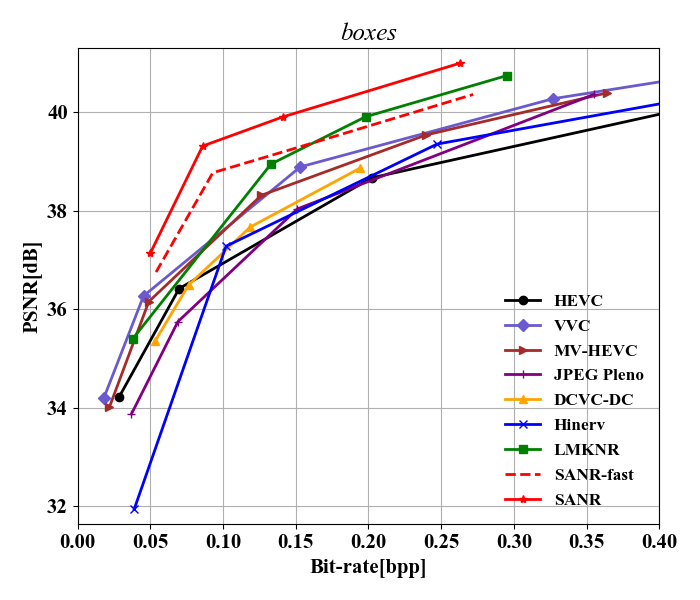} &
    \includegraphics[width=0.25\textwidth]{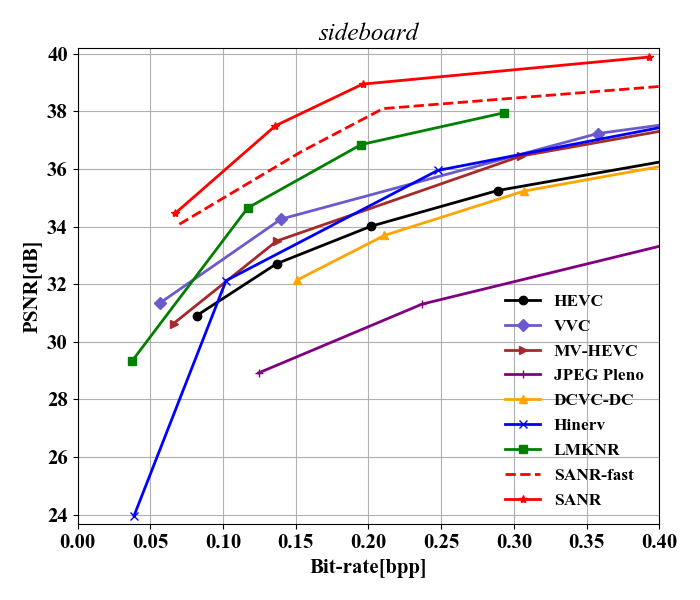} &
    \includegraphics[width=0.25\textwidth]{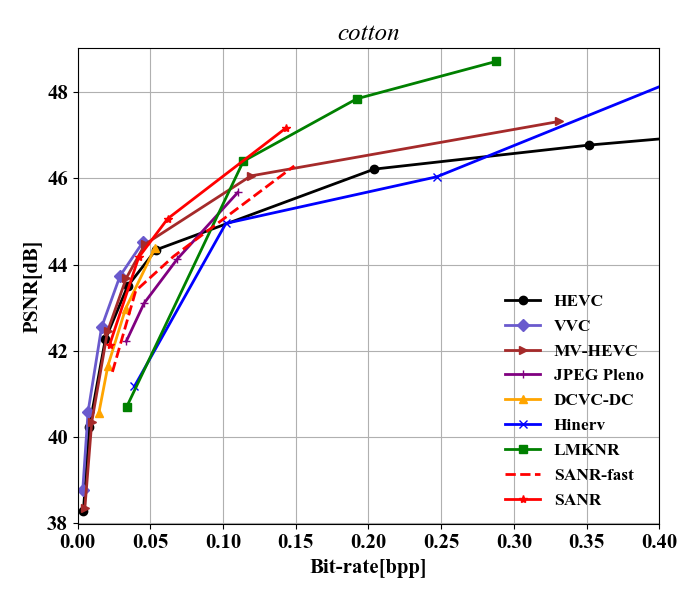} &
    \includegraphics[width=0.25\textwidth]{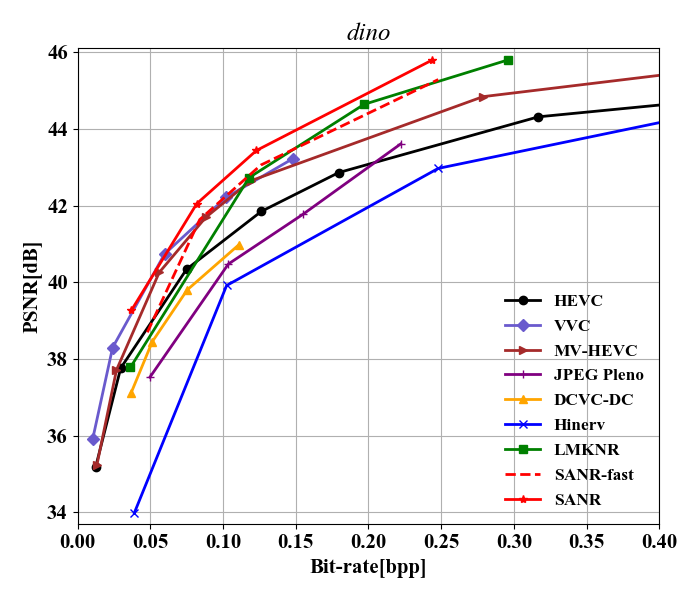}

  \end{tabular}
\end{tabular}

\caption{Rate-distortion performance of SANR and other methods.}
    \label{fig:RD curve}
\end{figure*}

In addition to the EPFL dataset, we utilize the HCI dataset and choose four synthetic scenes: \textit{boxes}, \textit{sideboard}, \textit{cotton}, and \textit{dino}. These scenes are rendered by Blender, a 3D graphics software, with a spatial resolution of 512 × 512 and an angular resolution of 9 × 9. The synthetic scenes simulate light field images captured by camera arrays, offering a lower noise level and a more comprehensive baseline than real-world light field images. This allows us to assess the performance of SANR on light field images with a more extensive disparity range, contributing to a more comprehensive evaluation.

\begin{figure*}[h]

\centering

\begin{tabular}{@{} c c @{}}
  \begin{tabular}{@{} c @{}}
    \rotatebox{90}{\textit{Bikes}} \\
  \end{tabular}
&
  \begin{tabular}{@{} c @{} c @{} c @{} c @{} c @{} c @{} c @{} c @{}}

    HEVC & MV-HEVC & VVC & DCVC-DC & HiNeRV & JPEG-P & LMKNR & SANR \\
    
    \includegraphics[width=0.1\textwidth]{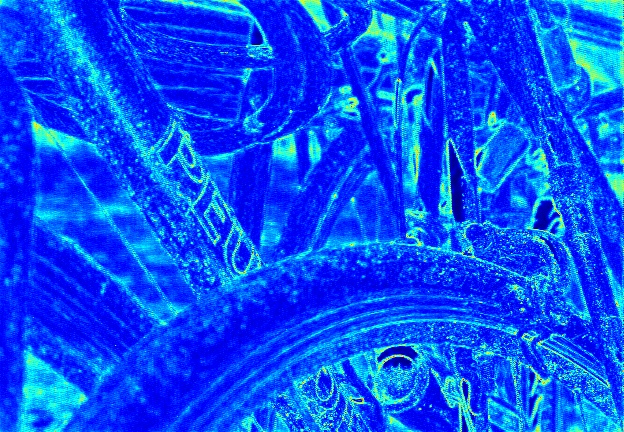} &
    \includegraphics[width=0.1\textwidth]{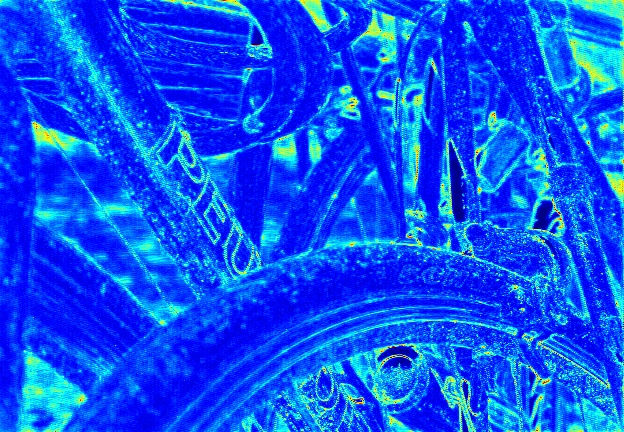} &
    \includegraphics[width=0.1\textwidth]{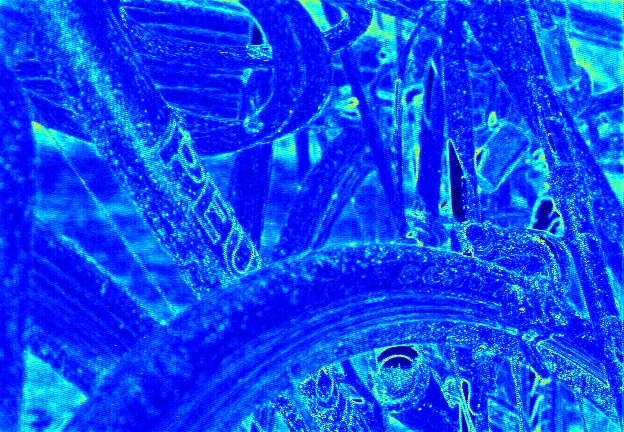} &
    \includegraphics[width=0.1\textwidth]{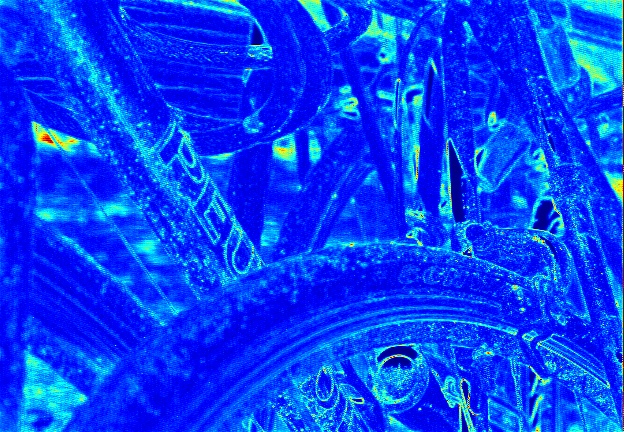} &
    \includegraphics[width=0.1\textwidth]{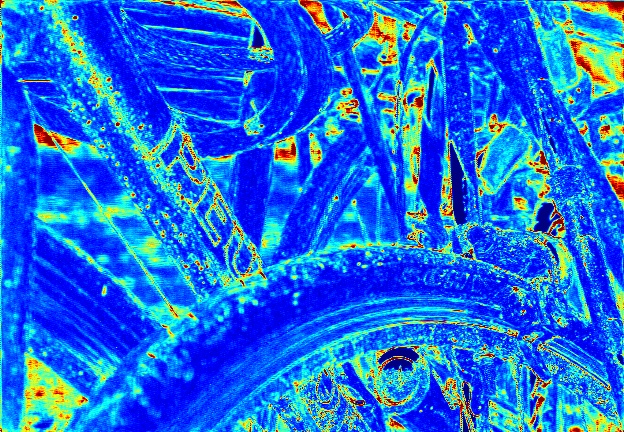} &
    \includegraphics[width=0.1\textwidth]{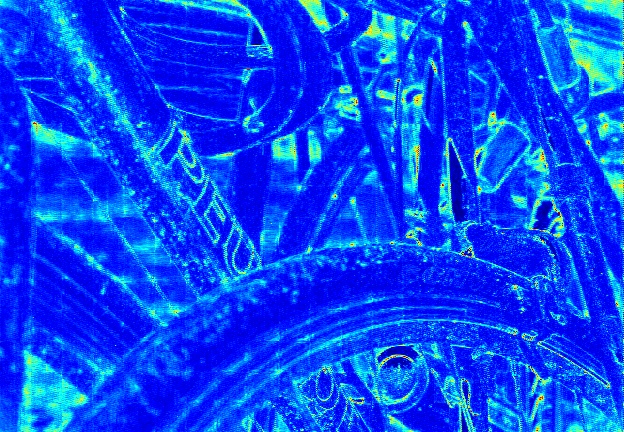} &
    \includegraphics[width=0.1\textwidth]{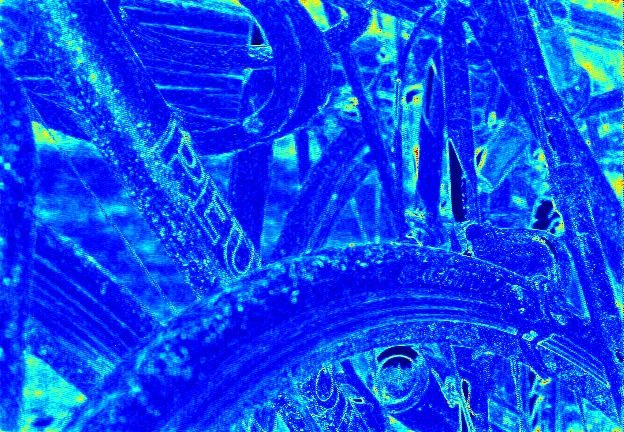} &
    \includegraphics[width=0.1\textwidth]{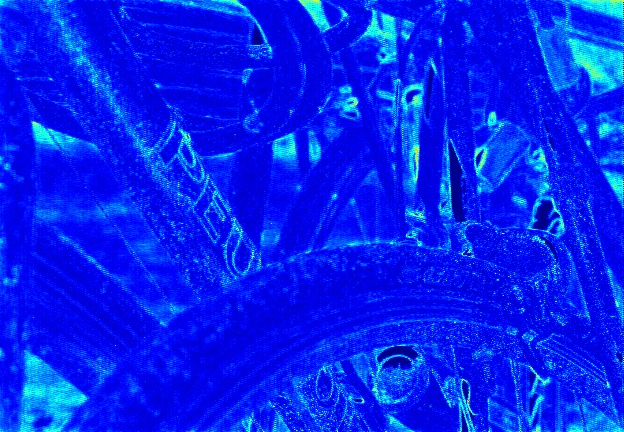} \\

    \includegraphics[width=0.1\textwidth]{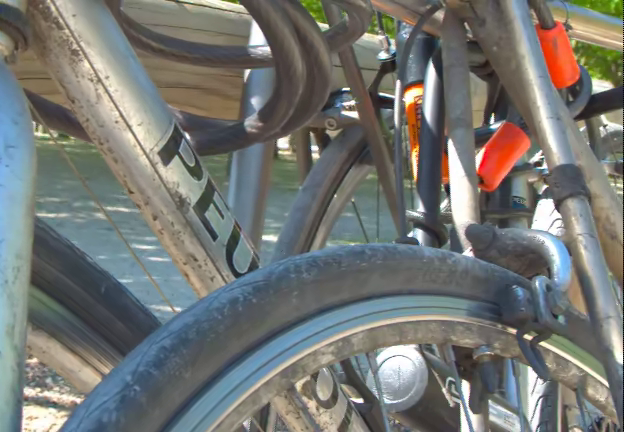} &
    \includegraphics[width=0.1\textwidth]{Fig/visual/HEVC/bikes.png} &
    \includegraphics[width=0.1\textwidth]{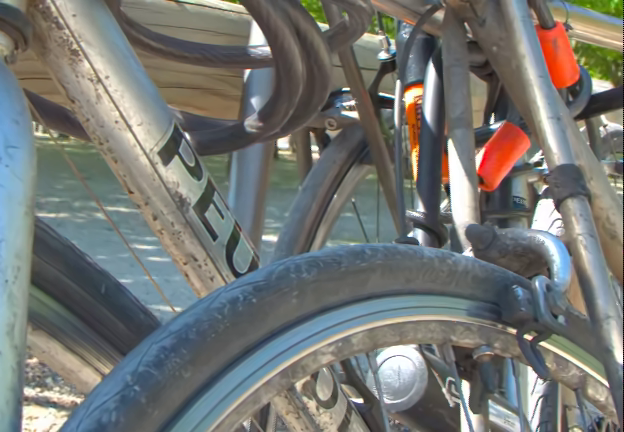} &
    \includegraphics[width=0.1\textwidth]{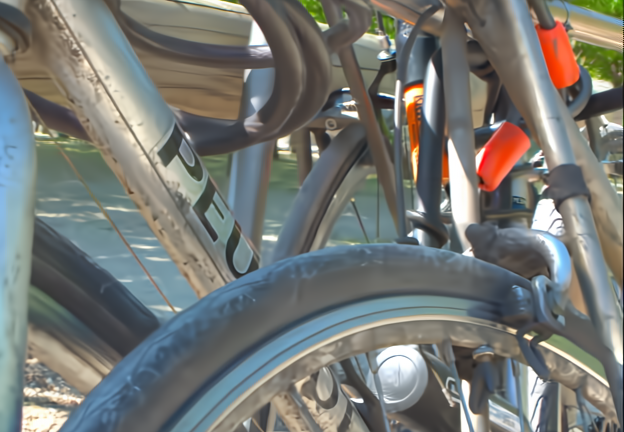} &
    \includegraphics[width=0.1\textwidth]{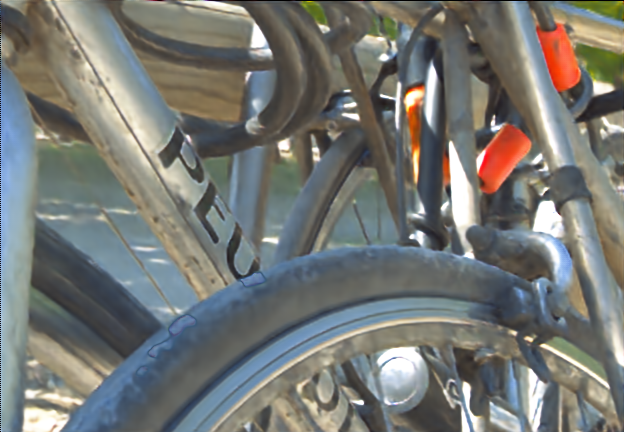} &
    \includegraphics[width=0.1\textwidth]{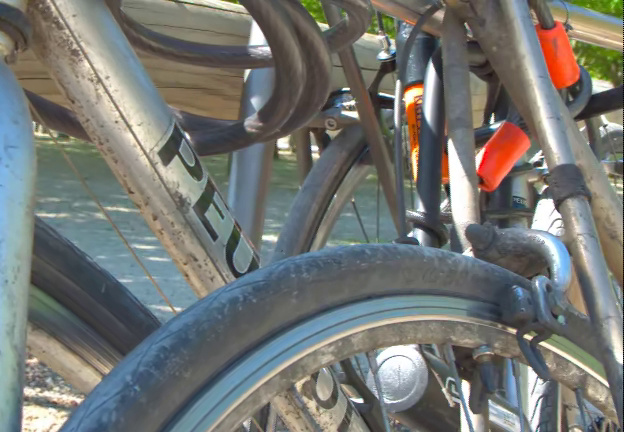} &
    \includegraphics[width=0.1\textwidth]{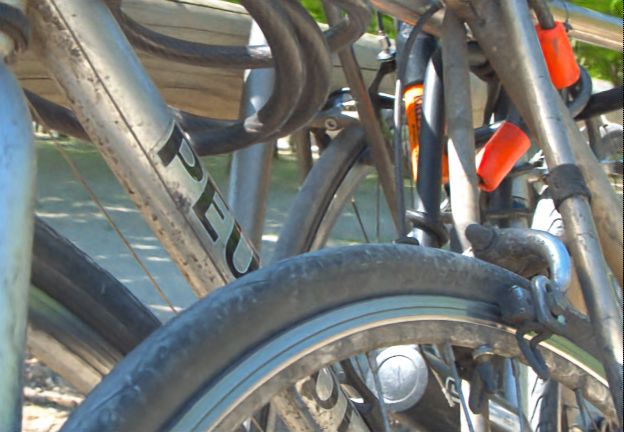} &
    \includegraphics[width=0.1\textwidth]{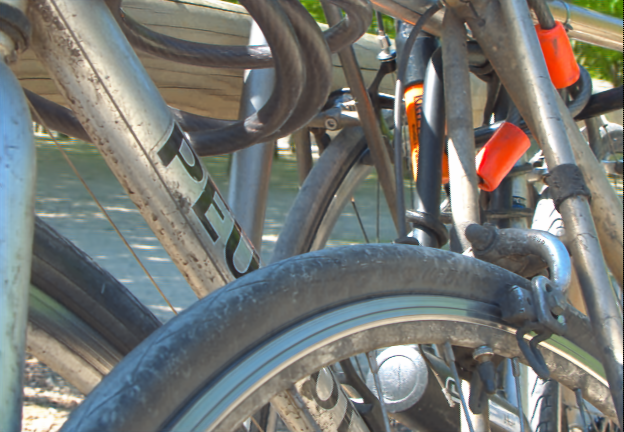} \\
    
     0.058 bpp & 0.047 bpp & 0.049 bpp &  0.091 bpp &  0.032 bpp &  0.060 bpp &  0.033 bpp &  0.04 bpp\\

      30.78 dB & 30.37 dB & 31.22 dB &  31.15 dB &  28.14 dB &  30.99 dB &  30.5 dB &  32.74 dB\\
  \end{tabular}
\end{tabular}

\begin{tabular}{@{} c c @{}}
  \begin{tabular}{@{} c @{}}
    \rotatebox{90}{\textit{sideboard}} \\
  \end{tabular}
&
  \begin{tabular}{@{} c @{} c @{} c @{} c @{} c @{} c @{} c @{} c @{}}

    \includegraphics[width=0.1\textwidth]{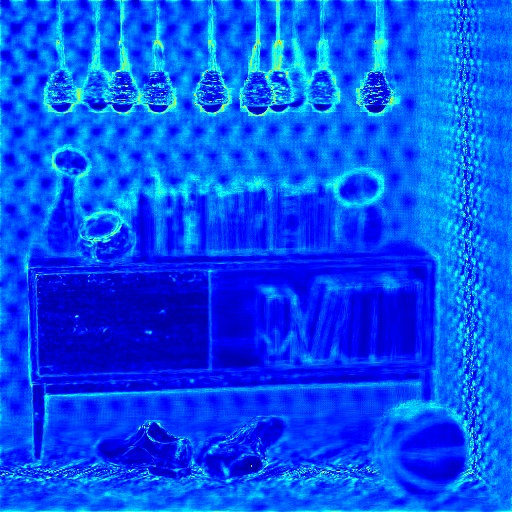} &
    \includegraphics[width=0.1\textwidth]{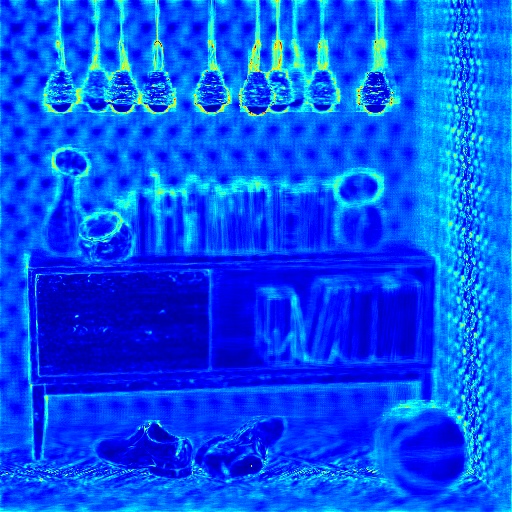} &
    \includegraphics[width=0.1\textwidth]{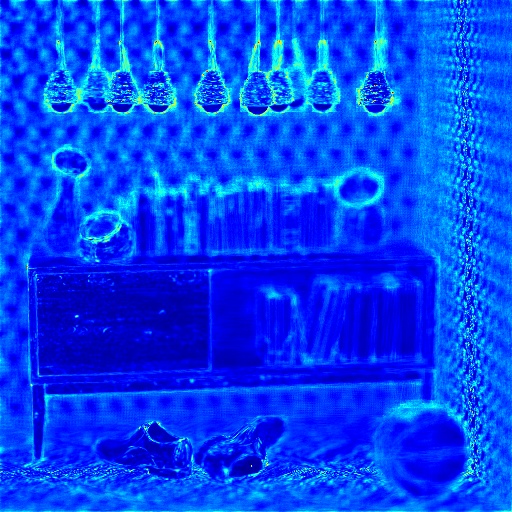} &
    \includegraphics[width=0.1\textwidth]{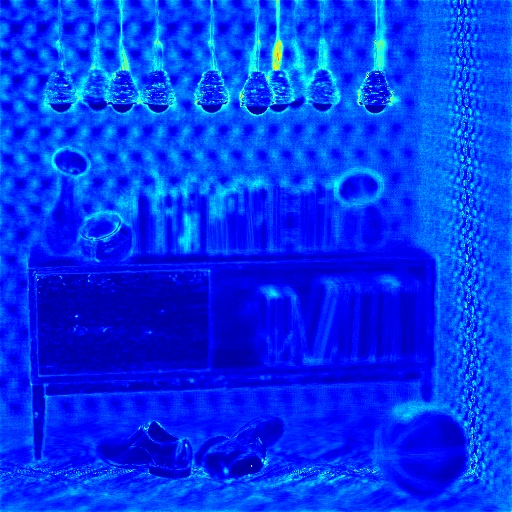} &
    \includegraphics[width=0.1\textwidth]{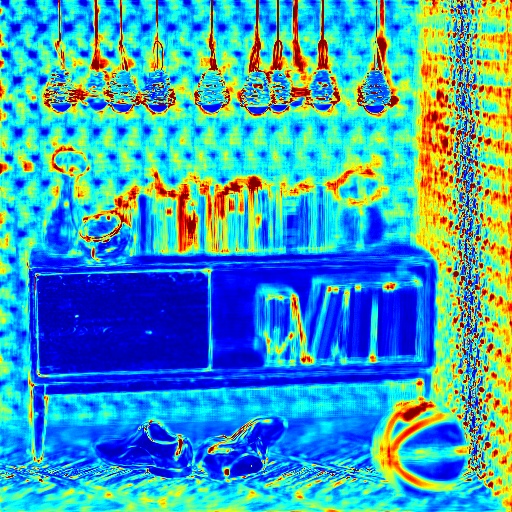} &
    \includegraphics[width=0.1\textwidth]{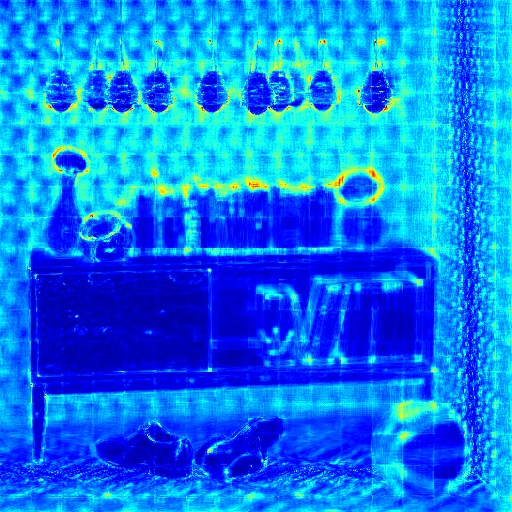} &
    \includegraphics[width=0.1\textwidth]{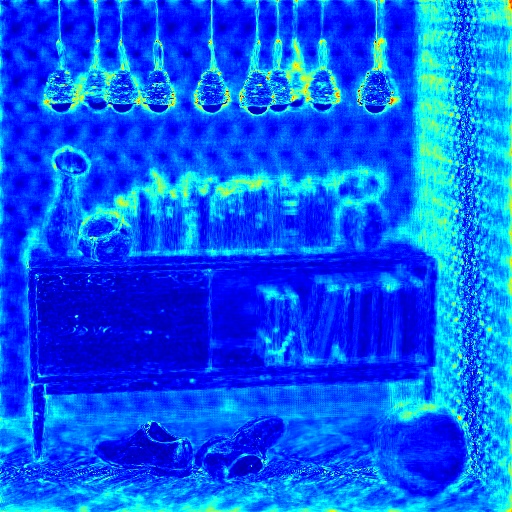} &
    \includegraphics[width=0.1\textwidth]{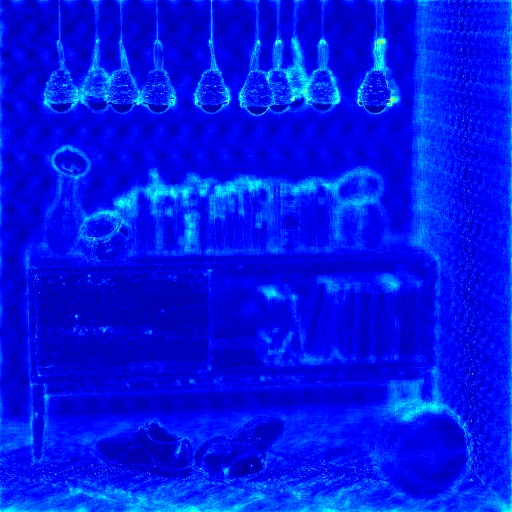}\\

    \includegraphics[width=0.1\textwidth]{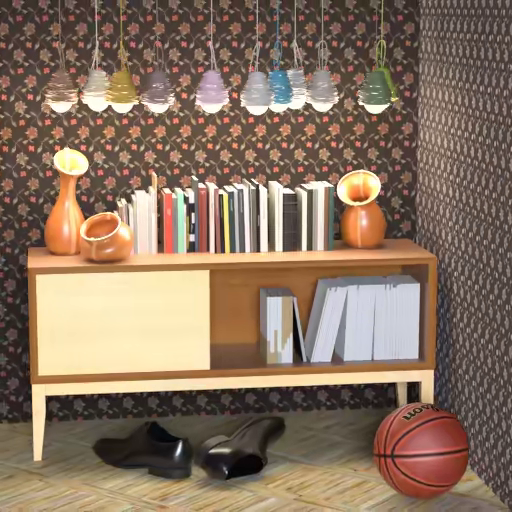} &
    \includegraphics[width=0.1\textwidth]{Fig/visual/HEVC/sideboard.png} &
    \includegraphics[width=0.1\textwidth]{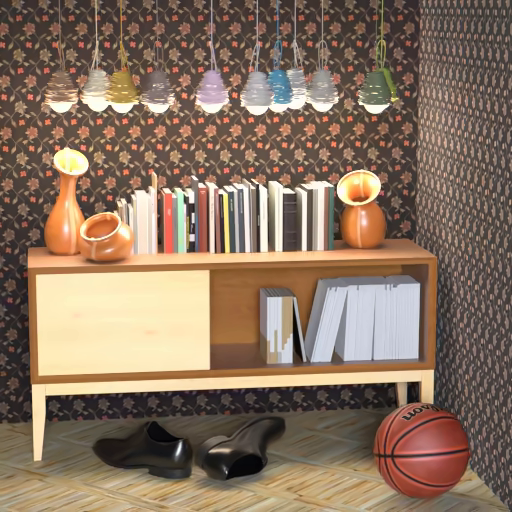} &
    \includegraphics[width=0.1\textwidth]{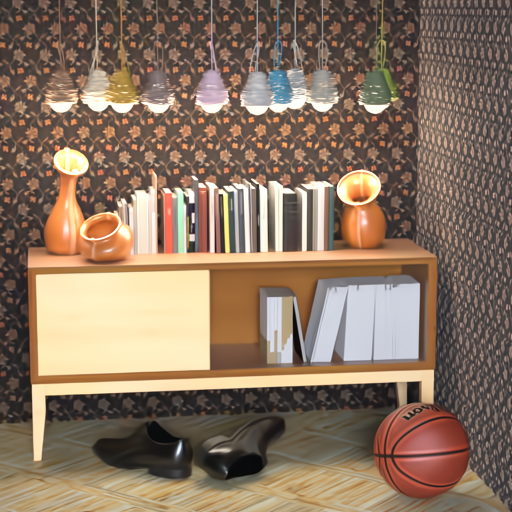} &
    \includegraphics[width=0.1\textwidth]{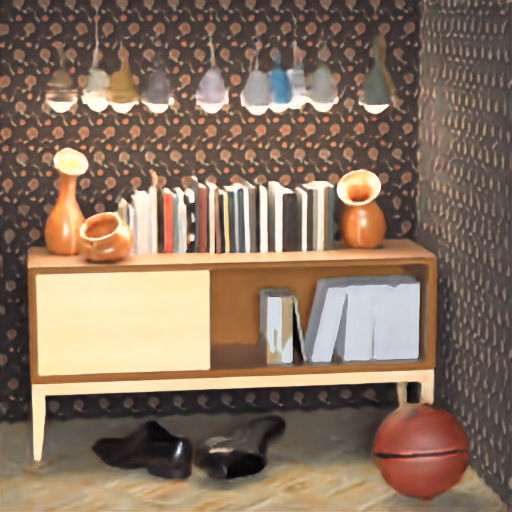} &
    \includegraphics[width=0.1\textwidth]{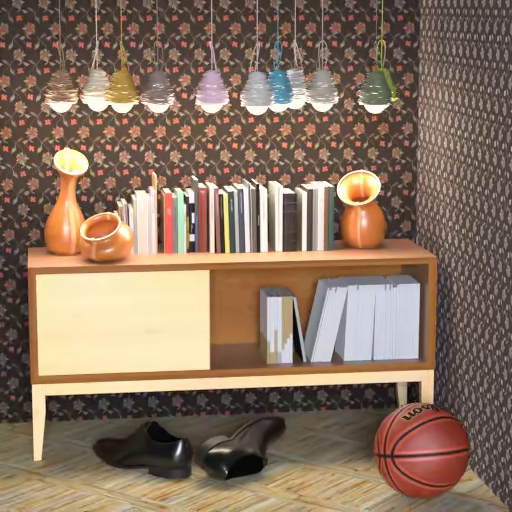} &
    \includegraphics[width=0.1\textwidth]{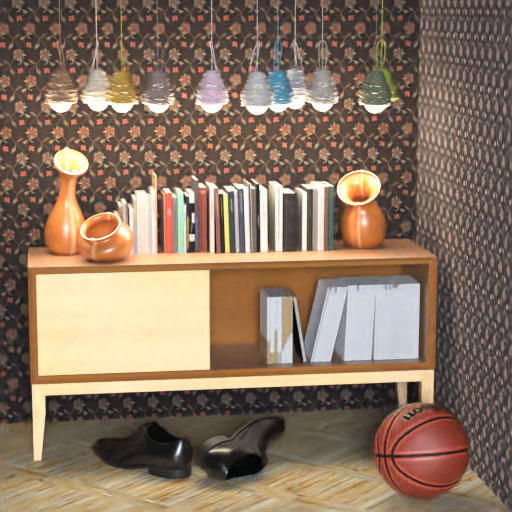} &
    \includegraphics[width=0.1\textwidth]{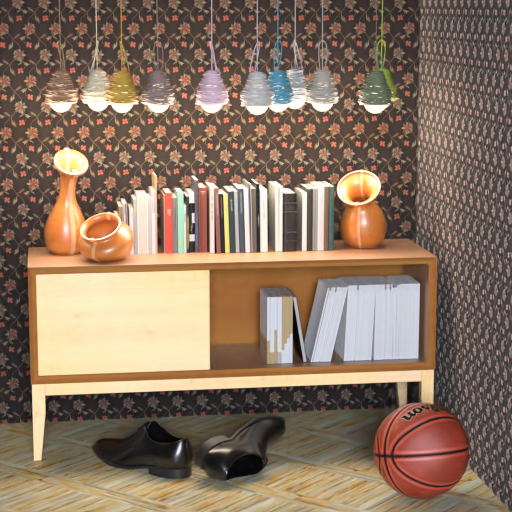} \\
    
     0.081 bpp & 0.066 bpp & 0.057 bpp & 0.151 bpp & 0.039 bpp & 0.124 bpp & 0.038 bpp & 0.067 bpp \\
       30.91 dB & 30.63 dB & 31.35 dB &  32.15 dB &  23.97 dB &  28.92 dB &  29.35 dB &  34.46 dB\\
  \end{tabular}
\end{tabular}

\caption{Average error heat map and the reconstructed central SAI of SANR and other methods at similar bpp. The error heat map gives the average absolute error of all SAIs in the light field image.}
\label{fig:avg error}

\end{figure*}

\subsubsection{Method Configurations}
We compare SANR against several state-of-the-art light field compression methods, spanning traditional, learned, and INR-based approaches:

\begin{itemize}
    \item \textbf{Traditional video codecs}: We treat the light field as a pseudo-video sequence and apply HEVC~\cite{hevc} (HM) and VVC~\cite{vvc} (VTM) under low-delay-P configurations. For multi-view coding, we test MV-HEVC using ``baseCfg-3view.cfg".
    
    \item \textbf{Learned video codec}: DCVC-DC~\cite{dcvcdc} is evaluated to encode the pseudo-video sequence of the light field.

    \item \textbf{INR-based video compression}: HiNeRV~\cite{kwan2023hinerv} is adapted to compress the pseudo-video sequence of light fields, with channel settings $\{30, 90, 160, 220\}$ for varying bitrates.

    \item \textbf{Light field-specific codec}: JPEG-Pleno~\cite{ebrahimi2016jpeg} is tested using the MuLE~\cite{MuLE} verification model.

    \item \textbf{INR-based light field method}: LKMNR~\cite{shi2023learning}, a modulated CNN-based approach, controls bitrate by varying the spatial kernel channel count. We adopt their default configuration with channel numbers $\{48, 93, 123, 153\}$ and a shared base tensor rank of 6.

    \item \textbf{SANR (Ours)}: Bitrate is controlled by adjusting both the spatial kernel channel count $C_S$ and the Lagrange multiplier $\lambda$ in the rate-distortion loss. The horizontal and vertical kernel channels are fixed to 1. The shared base tensor rank $r$ is set to 6, and the latent scene code channel number $C_l$ is set to 10. We evaluate four configurations with $C_S = \{48, 93, 123, 163\}$ and corresponding $\lambda$ values: $\{0.01, 0.005, 0.001, 0.0005\}$ for the EPFL dataset and $\{0.005, 0.001, 0.0005, 0.0001\}$ for the HCI dataset.
\end{itemize}

This diverse set of baselines enables a thorough evaluation of SANR across different compression paradigms and data characteristics.

\begin{figure*}[h]
    \centering
\begin{tabular}{@{} c @{}}

  \begin{tabular}{@{} c @{} c @{} c @{} c @{} c @{} c @{}}

    \includegraphics[width=0.25\textwidth]{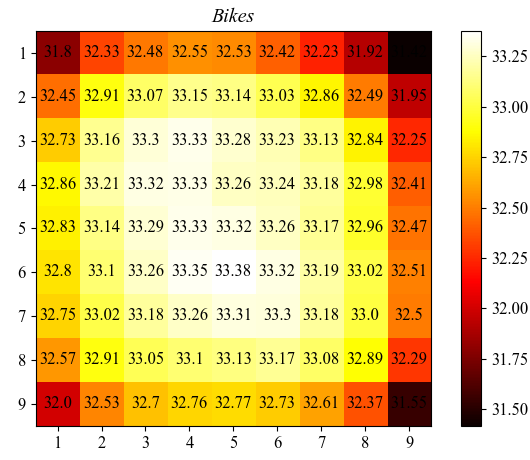} &
    \includegraphics[width=0.25\textwidth]{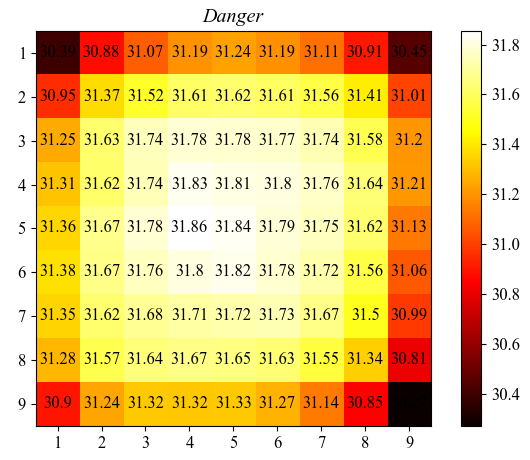} &
    \includegraphics[width=0.25\textwidth]{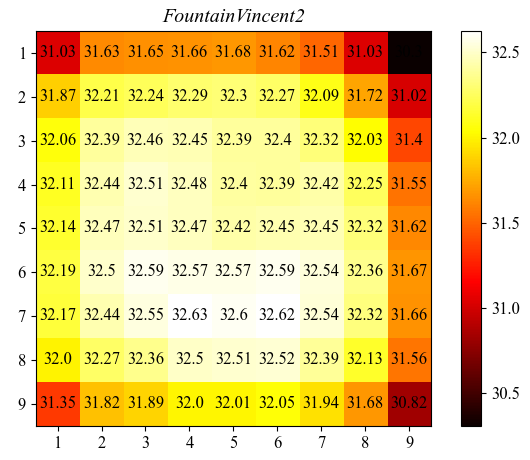} &
    \includegraphics[width=0.25\textwidth]{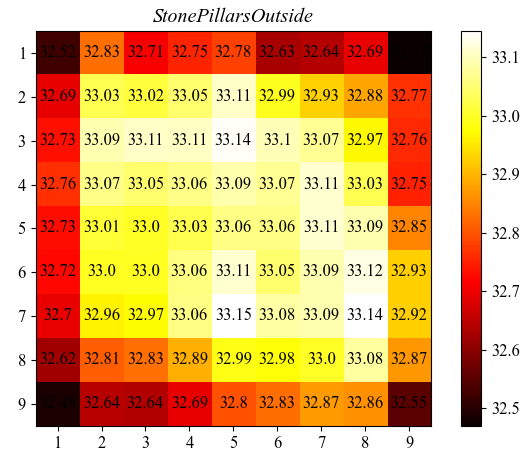}
    \\
    
    \includegraphics[width=0.25\textwidth]{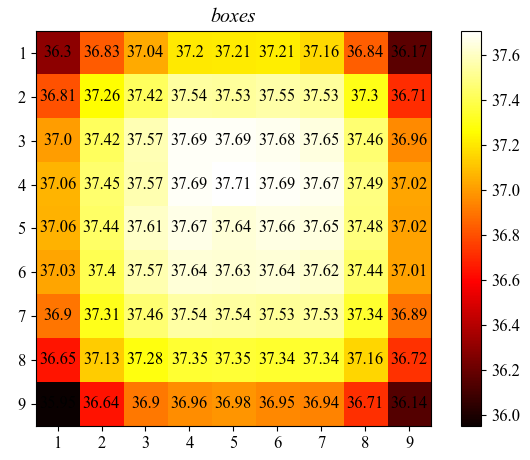} &
    \includegraphics[width=0.25\textwidth]{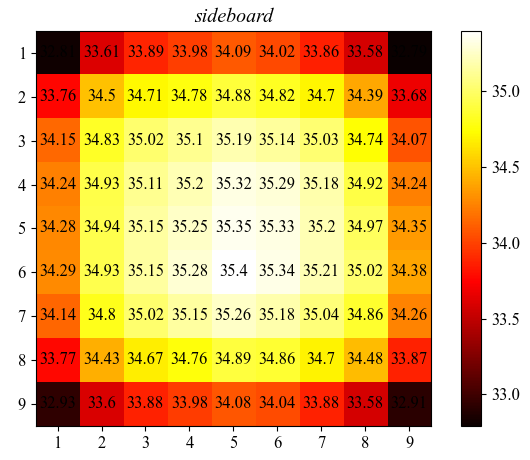} &
    \includegraphics[width=0.25\textwidth]{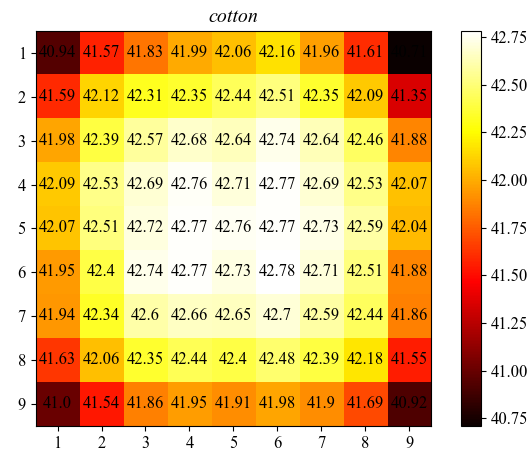} &
    \includegraphics[width=0.25\textwidth]{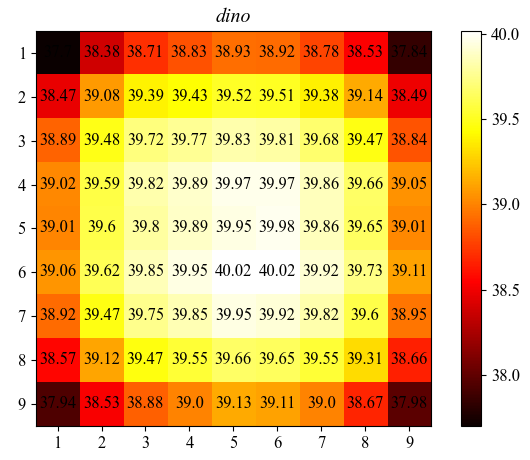}

  \end{tabular}
\end{tabular}

\caption{This figure gives the PSNR value of each SAI in the decoded light field image from SANR at the lowest bpp. It demonstrates consistency in the quality of central view reconstruction. }
    \label{fig:consistance}
\end{figure*}

\subsection{Performance Analysis}
In this section, we evaluate the performance of different methods through rate-distortion analysis and visual quality comparisons.

\subsubsection{Rate-Distortion Performance Comparison}
The rate-distortion curves are presented in Fig.\ref{fig:RD curve}, where the image quality is assessed in terms of PSNR and the bit rate is measured in bits per pixel (bpp). The rate-distortion curves indicate that SANR achieved the highest light field image quality among all comparison methods at the full bit rate range. In particular, SANR outperforms LMKNR on all light field images, which verifies the efficiency of the proposed hierarchical scene modeling block and the designed end-to-end rate-distortion optimization strategy for INR-based light field image compression. We also find that for video compression-based light field image compression methods, HiNeRV performs much better than HEVC, VVC, and DCVC-DC, especially at high bit rate. This result shows that the INR-based light field image compression scheme can better utilize the interrelationships between SAIs compared to the pseudo video sequence-based light field image compression methods and achieve more efficient compression performance. 

Table \ref{tab:bd rate} presents the BD-rate and BD-PSNR gains compared to the HEVC baseline for quantitative results. SANR shows consistent BD-rate gain on all light field images, with an average save of 65.62\% BD-rate gain against HEVC. Although VVC shows the least performance improvement compared to HEVC in Fig.\ref{fig:RD curve}, its average BD-rate save remains 25.77\%. This phenomenon indicates that the efficient utilization of inter-view relationships designed by other light field image compression methods significantly improves the performance of light field image compression. The gain of VVC compared to HEVC mainly comes from its ability to capture the connections between serial sequences. The continued superiority of other light field image compression methods over VVC is attributed to their utilization of non-serialized viewpoint relationships between light field images.

\subsubsection{Visual Comparison} To provide a visual comparison, we present average reconstruction error maps and the reconstructed central SAI at similar bit rates of different light field image compression methods in Fig.\ref{fig:avg error}. The error maps from all viewpoints reveal that SANR generates fewer errors and has a better reconstruction ability for texture in the foreground and background. For \textit{sideboard}, other methods make many more errors on the background wall, but SANR is good at compressing the background. For such periodic and detailed textures, as analyzed and shown in Fig.\ref{fig:scene visual}, SANR's small-scale latent scene codes play an important role in preserving this information. The large-scale latent scene codes contribute to reconstructing foreground objects, such as the lighting and basketball in the \textit{sideboard}. The average error maps can prove the effectiveness of the hierarchical scene modeling block for light field image compression.

\subsubsection{ Consistency Across Views} In Fig.\ref{fig:consistance}, we visualize the variation of PSNR values for each sub-aperture view of the reconstructed light field images at the lowest bpp of SANR. The central views show a consistent reconstruction quality. However, along the boundaries of the view array, the PSNR is worse than the central views, especially for the corner views. The boundary views lack some adjacent references and have worse PSNR than the central ones.

Overall, the results from rate-distortion analysis and visual comparisons demonstrate that SANR performs superiorly to other state-of-the-art methods across various evaluation metrics and scenarios.

\subsection{Encoding Complexity}
\label{sec:encoding_complexity}

Encoding complexity remains a universal limitation of INR‐based methods. We evaluate encoding time on the EPFL \textit{Bikes} sequences at low ($\approx$0.04 bpp) and high ($\approx$0.3 bpp) bitrates, all measured on a single NVIDIA V100 GPU. Tab.\ref{encoding speed} shows the encoding time of different light field image compression methods. All INR-based methods, including HiNeRV, LMKNR, and our SANR, show a much slower encoding speed than other traditional and end-to-end methods.

\begin{table}[htbp]
\centering
\begin{tabular}{c|cc}
\hline
                  & \textit{Bikes} (0.04 bpp) & \textit{Bikes} (0.3 bpp) \\ \hline
HEVC              &  2.98 min             & 5.26 min               \\ \hline
VVC               &  12.94 min             & 43.79 min                     \\ \hline
MV-HEVC           & 2.90 min             & 5.63 min                     \\ \hline
JPEG-P            & 0.38 min             & 0.58 min                      \\ \hline
DCVC-DC           & 0.10 min             & 0.11 min                     \\ \hline
HiNeRV            & 79.76 min             & 132.43 min                     \\ \hline
LMKNR             & 346.26 min             & 1250.73 min                      \\ \hline
SANR              & 1132.66 min             & 2031.58 min                      \\ \hline
SANR-fast              & 102.37 min             & 198.41 min                      \\ \hline
\end{tabular}

\caption{The encoding time of each light field image compression method on \textit{Bikes} sequences.}
\label{encoding speed}
\end{table}

To alleviate high encoding complexity, we propose an accelerated variant SANR-fast using DeepSpeed and reduced training epochs (6 regular, 1 SGA). As shown in Tab.\ref{encoding speed}, it saves encoding time by 90\% compared to SANR. Although rate-distortion performance drops compared to the SANR as illustrated in Fig.\ref{fig:RD curve}, it still outperforms other INR-based methods and significantly surpasses HiNeRV at similar encoding time. While slower than traditional or end-to-end approaches, its superior compression efficiency makes it suitable for storage-critical applications like document archiving, where encoding time can be traded for quality.

\begin{table}[htbp]
\centering
\begin{tabular}{c|cc}
\hline
                  & \textit{Bikes} (0.04 bpp) & \textit{Bikes} (0.3 bpp) \\ \hline
HEVC              &  0.666s/-             & 1.41s/-               \\ \hline
VVC               &  0.933s/-             & 1.808s/-                     \\ \hline
MV-HEVC           & 0.57s/-             & 1.095s/-                     \\ \hline
JPEG-P            & 4.1s/-             & 4.2s/-                      \\ \hline
DCVC-DC           & 5.44s/858.67             & 5.62s/858.67                     \\ \hline
HiNeRV            & 0.902s/0.016             & 1.059s/0.909                     \\ \hline
LMKNR             & 0.98s/0.38             & 1.85s/4.07                      \\ \hline
SANR              & 1.12s/0.41             & 1.98s/4.14                      \\ \hline
\end{tabular}

\caption{The decoding complexity of light field image compression methods on \textit{Bikes} sequences.}
\label{decode complexity}
\end{table}

\subsection{Decoding Complexity}
To thoroughly assess the practicality of SANR during deployment, we analyze its decoding complexity concerning runtime and kMAC per pixel, and compare it to other light field image codecs, including HEVC \cite{hevc}, VVC \cite{vvc}, MV-HEVC, JPEG-Pleno \cite{ebrahimi2016jpeg}, DCVC-DC \cite{dcvcdc}, HiNeRV \cite{kwan2023hinerv}, and LMKNR \cite{shi2023learning}. We test all learned methods on a V100 GPU and traditional methods on an AMD Ryzen 9 4900 CPU.

Tab.\ref{decode complexity} presents the decoding time and kMAC/pixel for \textit{Bikes} sequences at around 0.04 bpp and around 0.3 bpp. Conventional video codecs including HEVC, VVC, and MV-HEVC, demonstrate low latency. The decoding speed of JPEG Pleno, which is specifically designed for light field image compression, is slightly slower, around 4 seconds for decoding. Neural video codecs like DCVC-DC incur higher compute complexity about 858.67 kMAC/pixel but maintain reasonable runtime. For INR-based methods including SANR also show a fast decoding speed similar to traditional methods and these methods also show a much lower compute complexity than DCVC-DC.

\begin{figure}[h]
    \centering
    \includegraphics[width=0.8\linewidth]{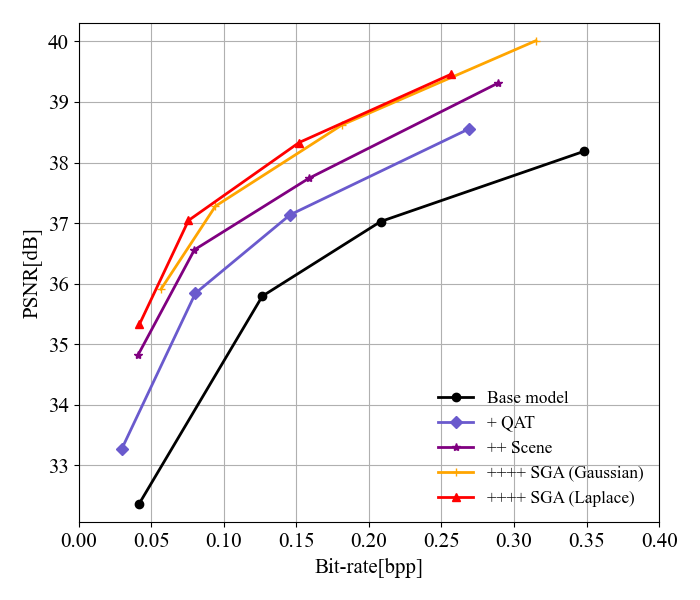}
    \caption{The average rate-distortion performance of our ablation studies among the eight sequences in the EPFL and HCI datasets.}
    \label{fig:ablation}
\end{figure}

\subsection{Ablation Study}
Here, we provide the ablation study using SANR. First, we train a base model without latent scene codes and conduct an 8-bit quantization on the trained base model. Then, we add the QAT technique, scene codes, and SGA to the base model. Herein, the +QAT, ++Scene, ++++SGA (Gaussiaan), ++++SGA (Laplace) correspond to based model+QAT, base model+QAT+scene codes, base model+QAT+scene codes, based model+QAT+scene codes+SGA with Gaussian entropy model, based model+QAT+scene codes+SGA with Laplace entropy model. The ablation study results are illustrated in Fig.\ref{fig:ablation}.

\subsubsection{QAT rate-distortion optimization}

In Fig.\ref{fig:ablation}, SANR performs much better with the rate-distortion optimization than the base model. And SANR with the Gaussian entropy model shows a weaker performance than SANR with the Laplace model. These results show that the separation processing of fitting INR to the light field image and compression of the INR cannot achieve the optimal rate-distortion performance. With a suitable entropy model, utilising QAT with the rate-distortion loss function in Eqn.(8) can solve the problem and improve the rate-distortion performance.

\subsubsection{Impact of latent scene codes}
In Fig.\ref{fig:ablation}, we show that the scene modeling block with the latent scene codes can get a better rate-distortion performance than the base model with QAT. This result demonstrates the effectiveness of the proposed hierarchical scene modeling block with the latent scene codes. As shown in Fig.\ref{fig:scene visual}, the average error maps demonstrate that the large-scale scene code concerns the details of the objects in the foreground, and the small-scale scene code concerns the periodic and smooth texture in the foreground and background.

\subsubsection{Stochastic Gumbel Annealing}
Fig.\ref{fig:ablation} shows that finetuning our model by SGA with three epochs helps improve the rate-distortion performance efficiently.

\section{Conclusion}
This paper proposes a scene-aware neural representation for light field image compression with rate-distortion optimization. We mainly addressed two problems for existing INR-based light field image compression: the lack of intrinsic scene features and the separation of light field reconstruction and compression. We proposed a hierarchical scene modeling block with latent scene codes to capture the intrinsic scene features. Experiments verified that the designed latent scene codes can capture texture details effectively. We proposed a QAT scheme with a rate-distortion loss function to train the model to achieve end-to-end rate-distortion optimization. Extensive experiments proved that SANR achieved state-of-the-art rate-distortion performance on widely utilized light field image databases.

\bibliographystyle{IEEEtran}
\bibliography{ref}

\begin{thebibliography}{10}
\providecommand{\url}[1]{#1}
\csname url@samestyle\endcsname
\providecommand{\newblock}{\relax}
\providecommand{\bibinfo}[2]{#2}
\providecommand{\BIBentrySTDinterwordspacing}{\spaceskip=0pt\relax}
\providecommand{\BIBentryALTinterwordstretchfactor}{4}
\providecommand{\BIBentryALTinterwordspacing}{\spaceskip=\fontdimen2\font plus
\BIBentryALTinterwordstretchfactor\fontdimen3\font minus \fontdimen4\font\relax}
\providecommand{\BIBforeignlanguage}[2]{{%
\expandafter\ifx\csname l@#1\endcsname\relax
\typeout{** WARNING: IEEEtran.bst: No hyphenation pattern has been}%
\typeout{** loaded for the language `#1'. Using the pattern for}%
\typeout{** the default language instead.}%
\else
\language=\csname l@#1\endcsname
\fi
#2}}
\providecommand{\BIBdecl}{\relax}
\BIBdecl

\bibitem{lf2016}
I.~Ihrke, J.~Restrepo, and L.~Mignard-Debise, ``Principles of light field imaging: Briefly revisiting 25 years of research,'' \emph{IEEE Signal Processing Magazine}, vol.~33, no.~5, pp. 59--69, 2016.

\bibitem{mazhan17}
\BIBentryALTinterwordspacing
Y.~Zhao, T.~Yue, L.~Chen, H.~Wang, Z.~Ma, D.~J. Brady, and X.~Cao, ``Heterogeneous camera array for multispectral light field imaging,'' \emph{Opt. Express}, vol.~25, no.~13, pp. 14\,008--14\,022, Jun 2017. [Online]. Available: \url{https://opg.optica.org/oe/abstract.cfm?URI=oe-25-13-14008}
\BIBentrySTDinterwordspacing

\bibitem{levoy2023light}
M.~Levoy and P.~Hanrahan, ``Light field rendering,'' in \emph{Seminal Graphics Papers: Pushing the Boundaries, Volume 2}, 2023, pp. 441--452.

\bibitem{jaykuo}
Y.~J. Jeong, J.-H. Lee, Y.~H. Cho, D.~Nam, D.-S. Park, and C.-C.~J. Kuo, ``Efficient light-field rendering using depth maps for 100-mpixel multi-projection 3d display,'' \emph{Journal of Display Technology}, vol.~11, no.~10, pp. 792--799, 2015.

\bibitem{cai2018ray}
Z.~Cai, X.~Liu, X.~Peng, and B.~Z. Gao, ``Ray calibration and phase mapping for structured-light-field 3d reconstruction,'' \emph{Optics Express}, vol.~26, no.~6, pp. 7598--7613, 2018.

\bibitem{wang2015occlusion}
T.-C. Wang, A.~A. Efros, and R.~Ramamoorthi, ``Occlusion-aware depth estimation using light-field cameras,'' in \emph{Proceedings of the IEEE international conference on computer vision}, 2015, pp. 3487--3495.

\bibitem{jeon2019pami}
H.-G. Jeon, J.~Park, G.~Choe, J.~Park, Y.~Bok, Y.-W. Tai, and I.~S. Kweon, ``Depth from a light field image with learning-based matching costs,'' \emph{IEEE Transactions on Pattern Analysis and Machine Intelligence}, vol.~41, no.~2, pp. 297--310, 2019.

\bibitem{mishiba2020fast}
K.~Mishiba, ``Fast depth estimation for light field cameras,'' \emph{IEEE Transactions on Image Processing}, vol.~29, pp. 4232--4242, 2020.

\bibitem{yu2017light}
J.~Yu, ``A light-field journey to virtual reality,'' \emph{IEEE MultiMedia}, vol.~24, no.~2, pp. 104--112, 2017.

\bibitem{overbeck2018system}
R.~S. Overbeck, D.~Erickson, D.~Evangelakos, M.~Pharr, and P.~Debevec, ``A system for acquiring, processing, and rendering panoramic light field stills for virtual reality,'' \emph{ACM Transactions on Graphics (TOG)}, vol.~37, no.~6, pp. 1--15, 2018.

\bibitem{zhangqi2023vr}
Q.~Zhang, J.~Wei, S.~Wang, S.~Ma, and W.~Gao, ``Realvr: Efficient, economical, and quality-of- experience-driven vr video system based on mpeg omaf,'' \emph{IEEE Transactions on Multimedia}, vol.~25, pp. 5386--5399, 2023.

\bibitem{1993JPEG}
W.~B. Pennebaker and J.~L. Mitchell, ``{JPEG still image data compression standard},'' \emph{Van Nostrand Reinhold}, 1993.

\bibitem{2015BPG}
F.~Bellard, ``{BPG Image Format},'' 2015.

\bibitem{monteiro2016light}
R.~Monteiro, L.~Lucas, C.~Conti, P.~Nunes, N.~Rodrigues, S.~Faria, C.~Pagliari, E.~Da~Silva, and L.~Soares, ``{Light field HEVC-based image coding using locally linear embedding and self-similarity compensated prediction},'' in \emph{2016 IEEE International Conference on Multimedia \& Expo Workshops (ICMEW)}.\hskip 1em plus 0.5em minus 0.4em\relax IEEE, 2016, pp. 1--4.

\bibitem{hevc}
G.~J. Sullivan, J.-R. Ohm, W.-J. Han, and T.~Wiegand, ``{Overview of the high efficiency video coding (HEVC) standard},'' \emph{IEEE Transactions on circuits and systems for video technology}, vol.~22, no.~12, pp. 1649--1668, 2012.

\bibitem{astola2018wasp}
P.~Astola and I.~Tabus, ``Wasp: Hierarchical warping, merging, and sparse prediction for light field image compression,'' in \emph{2018 7th European Workshop on Visual Information Processing (EUVIP)}.\hskip 1em plus 0.5em minus 0.4em\relax IEEE, 2018, pp. 1--6.

\bibitem{alves2020jpeg}
G.~D.~O. Alves, M.~B. De~Carvalho, C.~L. Pagliari, P.~G. Freitas, I.~Seidel, M.~P. Pereira, C.~F.~S. Vieira, V.~Testoni, F.~Pereira, and E.~A. Da~Silva, ``{The JPEG Pleno light field coding standard 4D-transform mode: How to design an efficient 4D-native codec},'' \emph{IEEE Access}, vol.~8, pp. 170\,807--170\,829, 2020.

\bibitem{mildenhall2020nerf}
B.~Mildenhall, P.~Srinivasan, M.~Tancik, J.~Barron, R.~Ramamoorthi, and R.~Ng, ``Nerf: Representing scenes as neural radiance fields for view synthesis,'' in \emph{European conference on computer vision}, 2020.

\bibitem{dupont2021coin}
E.~Dupont, A.~Golinski, M.~Alizadeh, Y.~W. Teh, and A.~Doucet, ``Coin: Compression with implicit neural representations,'' in \emph{Neural Compression: From Information Theory to Applications--Workshop@ ICLR 2021}, 2021.

\bibitem{chen2021nerv}
H.~Chen, B.~He, H.~Wang, Y.~Ren, S.~N. Lim, and A.~Shrivastava, ``Nerv: Neural representations for videos,'' \emph{Advances in Neural Information Processing Systems}, vol.~34, pp. 21\,557--21\,568, 2021.

\bibitem{shi2022distilled}
J.~Shi and C.~Guillemot, ``Distilled low rank neural radiance field with quantization for light field compression,'' \emph{arXiv preprint arXiv:2208.00164}, 2022.

\bibitem{wang2022light}
H.~Wang, H.~Zhu, and Z.~Chen, ``Light field compression based on implicit neural representation,'' in \emph{2022 Picture Coding Symposium (PCS)}.\hskip 1em plus 0.5em minus 0.4em\relax IEEE, 2022, pp. 223--227.

\bibitem{jiang2022untrained}
X.~Jiang, J.~Shi, and C.~Guillemot, ``An untrained neural network prior for light field compression,'' \emph{IEEE Transactions on Image Processing}, vol.~31, pp. 6922--6936, 2022.

\bibitem{shi2023learning}
J.~Shi, Y.~Xu, and C.~Guillemot, ``Learning kernel-modulated neural representation for efficient light field compression,'' \emph{IEEE Transactions on Image Processing}, 2024.

\bibitem{dai2015lenselet}
F.~Dai, J.~Zhang, Y.~Ma, and Y.~Zhang, ``Lenselet image compression scheme based on subaperture images streaming,'' in \emph{2015 IEEE International Conference on Image Processing (ICIP)}.\hskip 1em plus 0.5em minus 0.4em\relax IEEE, 2015, pp. 4733--4737.

\bibitem{li2017pseudo}
L.~Li, Z.~Li, B.~Li, D.~Liu, and H.~Li, ``Pseudo-sequence-based 2-d hierarchical coding structure for light-field image compression,'' \emph{IEEE Journal of Selected Topics in Signal Processing}, vol.~11, no.~7, pp. 1107--1119, 2017.

\bibitem{ahmad2020shearlet}
W.~Ahmad, S.~Vagharshakyan, M.~Sj{\"o}str{\"o}m, A.~Gotchev, R.~Bregovic, and R.~Olsson, ``Shearlet transform-based light field compression under low bitrates,'' \emph{IEEE Transactions on Image Processing}, vol.~29, pp. 4269--4280, 2020.

\bibitem{rizkallah2019geometry}
M.~Rizkallah, X.~Su, T.~Maugey, and C.~Guillemot, ``Geometry-aware graph transforms for light field compact representation,'' \emph{IEEE Transactions on Image Processing}, vol.~29, pp. 602--616, 2019.

\bibitem{hou2023TMM}
Y.~Zhang, W.~Dai, Y.~Li, C.~Li, J.~Hou, J.~Zou, and H.~Xiong, ``Light field compression with graph learning and dictionary-guided sparse coding,'' \emph{IEEE Transactions on Multimedia}, vol.~25, pp. 3059--3072, 2023.

\bibitem{liu2022multi}
D.~Liu, Y.~Huang, Y.~Fang, Y.~Zuo, and P.~An, ``Multi-stream dense view reconstruction network for light field image compression,'' \emph{IEEE Transactions on Multimedia}, 2022.

\bibitem{zhang2022light}
Y.~Zhang, W.~Dai, Y.~Li, C.~Li, J.~Hou, J.~Zou, and H.~Xiong, ``Light field compression with graph learning and dictionary-guided sparse coding,'' \emph{IEEE Transactions on Multimedia}, 2022.

\bibitem{hu2021multiple}
X.~Hu, Y.~Pan, Y.~Wang, L.~Zhang, and S.~Shirmohammadi, ``Multiple description coding for best-effort delivery of light field video using gnn-based compression,'' \emph{IEEE Transactions on Multimedia}, 2021.

\bibitem{huang2023prediction}
X.~Huang, Y.~Chen, P.~An, and L.~Shen, ``Prediction-oriented disparity rectification model for geometry-based light field compression,'' \emph{IEEE Transactions on Broadcasting}, vol.~69, no.~1, pp. 62--74, 2023.

\bibitem{tong2022sadn}
K.~Tong, X.~Jin, C.~Wang, and F.~Jiang, ``Sadn: learned light field image compression with spatial-angular decorrelation,'' in \emph{ICASSP 2022-2022 IEEE International Conference on Acoustics, Speech and Signal Processing (ICASSP)}.\hskip 1em plus 0.5em minus 0.4em\relax IEEE, 2022, pp. 1870--1874.

\bibitem{stanley2007compositional}
K.~O. Stanley, ``Compositional pattern producing networks: A novel abstraction of development,'' \emph{Genetic programming and evolvable machines}, vol.~8, pp. 131--162, 2007.

\bibitem{sitzmann2020implicit}
V.~Sitzmann, J.~Martel, A.~Bergman, D.~Lindell, and G.~Wetzstein, ``Implicit neural representations with periodic activation functions,'' \emph{Advances in neural information processing systems}, vol.~33, pp. 7462--7473, 2020.

\bibitem{chen2021learning}
Y.~Chen, S.~Liu, and X.~Wang, ``Learning continuous image representation with local implicit image function,'' in \emph{Proceedings of the IEEE/CVF conference on computer vision and pattern recognition}, 2021, pp. 8628--8638.

\bibitem{zhang2022implicit}
K.~Zhang, D.~Zhu, X.~Min, and G.~Zhai, ``Implicit neural representation learning for hyperspectral image super-resolution,'' \emph{IEEE Transactions on Geoscience and Remote Sensing}, vol.~61, pp. 1--12, 2022.

\bibitem{gordon2023quantizing}
C.~Gordon, S.-F. Chng, L.~MacDonald, and S.~Lucey, ``On quantizing implicit neural representations,'' in \emph{IEEE/CVF Winter Conference on Applications of Computer Vision}, 2023, pp. 341--350.

\bibitem{strumpler2022implicit}
Y.~Str{\"u}mpler, J.~Postels, R.~Yang, L.~V. Gool, and F.~Tombari, ``Implicit neural representations for image compression,'' in \emph{Computer Vision--ECCV 2022: 17th European Conference, Tel Aviv, Israel, October 23--27, 2022, Proceedings, Part XXVI}.\hskip 1em plus 0.5em minus 0.4em\relax Springer, 2022, pp. 74--91.

\bibitem{li2022nerv}
Z.~Li, M.~Wang, H.~Pi, K.~Xu, J.~Mei, and Y.~Liu, ``E-nerv: Expedite neural video representation with disentangled spatial-temporal context,'' in \emph{European Conference on Computer Vision}.\hskip 1em plus 0.5em minus 0.4em\relax Springer, 2022, pp. 267--284.

\bibitem{chen2023hnerv}
H.~Chen, M.~Gwilliam, S.-N. Lim, and A.~Shrivastava, ``Hnerv: A hybrid neural representation for videos,'' in \emph{Proceedings of the IEEE/CVF Conference on Computer Vision and Pattern Recognition}, 2023, pp. 10\,270--10\,279.

\bibitem{kwan2023hinerv}
H.~M. Kwan, G.~Gao, F.~Zhang, A.~Gower, and D.~Bull, ``Hinerv: Video compression with hierarchical encoding-based neural representation,'' \emph{Advances in Neural Information Processing Systems}, vol.~36, 2024.

\bibitem{barron2021mip}
J.~T. Barron, B.~Mildenhall, M.~Tancik, P.~Hedman, R.~Martin-Brualla, and P.~P. Srinivasan, ``Mip-nerf: A multiscale representation for anti-aliasing neural radiance fields,'' in \emph{Proceedings of the IEEE/CVF International Conference on Computer Vision}, 2021, pp. 5855--5864.

\bibitem{muller2022instant}
T.~M{\"u}ller, A.~Evans, C.~Schied, and A.~Keller, ``Instant neural graphics primitives with a multiresolution hash encoding,'' \emph{ACM Transactions on Graphics (ToG)}, vol.~41, no.~4, pp. 1--15, 2022.

\bibitem{feng2021signet}
B.~Y. Feng and A.~Varshney, ``Signet: Efficient neural representation for light fields,'' in \emph{Proceedings of the IEEE/CVF International Conference on Computer Vision}, 2021, pp. 14\,224--14\,233.

\bibitem{feng2022neural}
------, ``Neural subspaces for light fields,'' \emph{IEEE Transactions on Visualization and Computer Graphics}, vol.~30, no.~3, pp. 1685--1695, 2022.

\bibitem{yang2020improving}
Y.~Yang, R.~Bamler, and S.~Mandt, ``Improving inference for neural image compression,'' \emph{Advances in Neural Information Processing Systems}, vol.~33, pp. 573--584, 2020.

\bibitem{witten1987arithmetic}
I.~H. Witten, R.~M. Neal, and J.~G. Cleary, ``Arithmetic coding for data compression,'' \emph{Communications of the ACM}, vol.~30, no.~6, pp. 520--540, 1987.

\bibitem{ste}
P.~Yin, J.~Lyu, S.~Zhang, S.~Osher, Y.~Qi, and J.~Xin, ``Understanding straight-through estimator in training activation quantized neural nets,'' in \emph{International Conference on Learning Representations}, 2019.

\bibitem{kingma2014adam}
D.~Kingma, ``Adam: a method for stochastic optimization,'' in \emph{Int Conf Learn Represent}, 2014.

\bibitem{rerabek2016new}
M.~Rerabek and T.~Ebrahimi, ``New light field image dataset,'' in \emph{8th International Conference on Quality of Multimedia Experience (QoMEX)}, no. CONF, 2016.

\bibitem{honauer2017dataset}
K.~Honauer, O.~Johannsen, D.~Kondermann, and B.~Goldluecke, ``A dataset and evaluation methodology for depth estimation on 4d light fields,'' in \emph{Computer Vision--ACCV 2016: 13th Asian Conference on Computer Vision, Taipei, Taiwan, November 20-24, 2016, Revised Selected Papers, Part III 13}.\hskip 1em plus 0.5em minus 0.4em\relax Springer, 2017, pp. 19--34.

\bibitem{vvc}
B.~Bross, Y.-K. Wang, Y.~Ye, S.~Liu, J.~Chen, G.~J. Sullivan, and J.-R. Ohm, ``{Overview of the versatile video coding (VVC) standard and its applications},'' \emph{IEEE Transactions on Circuits and Systems for Video Technology}, vol.~31, no.~10, pp. 3736--3764, 2021.

\bibitem{dcvcdc}
J.~Li, B.~Li, and Y.~Lu, ``Neural video compression with diverse contexts,'' in \emph{Proceedings of the IEEE/CVF conference on computer vision and pattern recognition}, 2023, pp. 22\,616--22\,626.

\bibitem{ebrahimi2016jpeg}
T.~Ebrahimi, S.~Foessel, F.~Pereira, and P.~Schelkens, ``{JPEG Pleno: Toward an efficient representation of visual reality},'' \emph{IEEE Multimedia}, vol.~23, no.~4, pp. 14--20, 2016.

\bibitem{MuLE}
M.~B. de~Carvalho, M.~P. Pereira, G.~Alves, E.~A. da~Silva, C.~L. Pagliari, F.~Pereira, and V.~Testoni, ``A 4d dct-based lenslet light field codec,'' in \emph{2018 25th IEEE International Conference on Image Processing (ICIP)}.\hskip 1em plus 0.5em minus 0.4em\relax IEEE, 2018, pp. 435--439.

\end{thebibliography}
\end{document}